# Hierarchical Structural Analysis Method for Complex Equation-Oriented Models

**Chao Wang [1], Li Wan [1], Tifan Xiong [1,*], Yuanlong Xie [1,2], Shuting Wang [1,2], Jianwan Ding [1] and Liping Chen [1]**

1 School of Mechanical Science and Engineering, Huazhong University of Science and Technology, Wuhan 430074, China; wangc@hust.edu.cn (C.W.); wanli@hust.edu.cn (L.W.); yuanlongxie@hust.edu.cn (Y.X.); wangst@hust.edu.cn (S.W.); dingjw@hust.edu.cn (J.D.); chenlp@hust.edu.cn (L.C.)
2 Guangdong Intelligent Robotics Institute, Dongguan, 523808, China
* Correspondence: xiongtf@hust.edu.cn

**Abstract:** Structural analysis is a method for verifying equation-oriented models in the design of industrial systems. Existing structural analysis methods need flattening of the hierarchical models into an equation system for analysis. However, the large-scale equations in complex models make structural analysis difficult. Aimed to address the issue, this study proposes a hierarchical structural analysis method by exploring the relationship between the singularities of the hierarchical equation-oriented model and its components. This method obtains the singularity of a hierarchical equation-oriented model by analyzing a dummy model constructed with the parts from the decomposing results of its components. Based on this, the structural singularity of a complex model can be obtained by layer-by-layer analysis according to their natural hierarchy. The hierarchical structural analysis method can reduce the equation scale in each analysis and achieve efficient structural analysis of very complex models. This method can be adaptively applied to nonlinear-algebraic and differential-algebraic equation models. The main algorithms, application cases and comparison with the existing methods are present in this paper. The complexity analysis results show the enhanced efficiency of the proposed method in the structural analysis of complex equation-oriented models. Compared with the existing methods, the time complexity of the proposed method is improved significantly.

## 1. Introduction

With the increasing complexity and scale of modern industrial products, model-based approaches have become essential in studying these products, being applied in areas such as design [1,2], simulation [3] and diagnosis [4–6]. The modular models in different abstraction levels enable collaborative work and component reuse and speed up product modeling with a V-shaped process, as shown in Figure 1. The requirements are hierarchically decomposed into the modeling constraints of subsystems and components in different abstraction levels. These models are constructed based on the predefined components or the models in the lower level, thereby forming a hierarchical structure. Equation-oriented models (EoMs) are often adopted to model multi-domain systems because of their convenience in modeling and ability to express physical characteristics [7]. Guided by the modeling purpose, hierarchical EoMs can abstract putative systems to predict states and behavior effectively [1]. Languages and tools for EoMs, such as Modelica [8], gPROMS [9], Dymola [10] and MWorks [11], have been widely investigated in engineering applications to express the static and dynamic characteristics of physical systems [7,12].

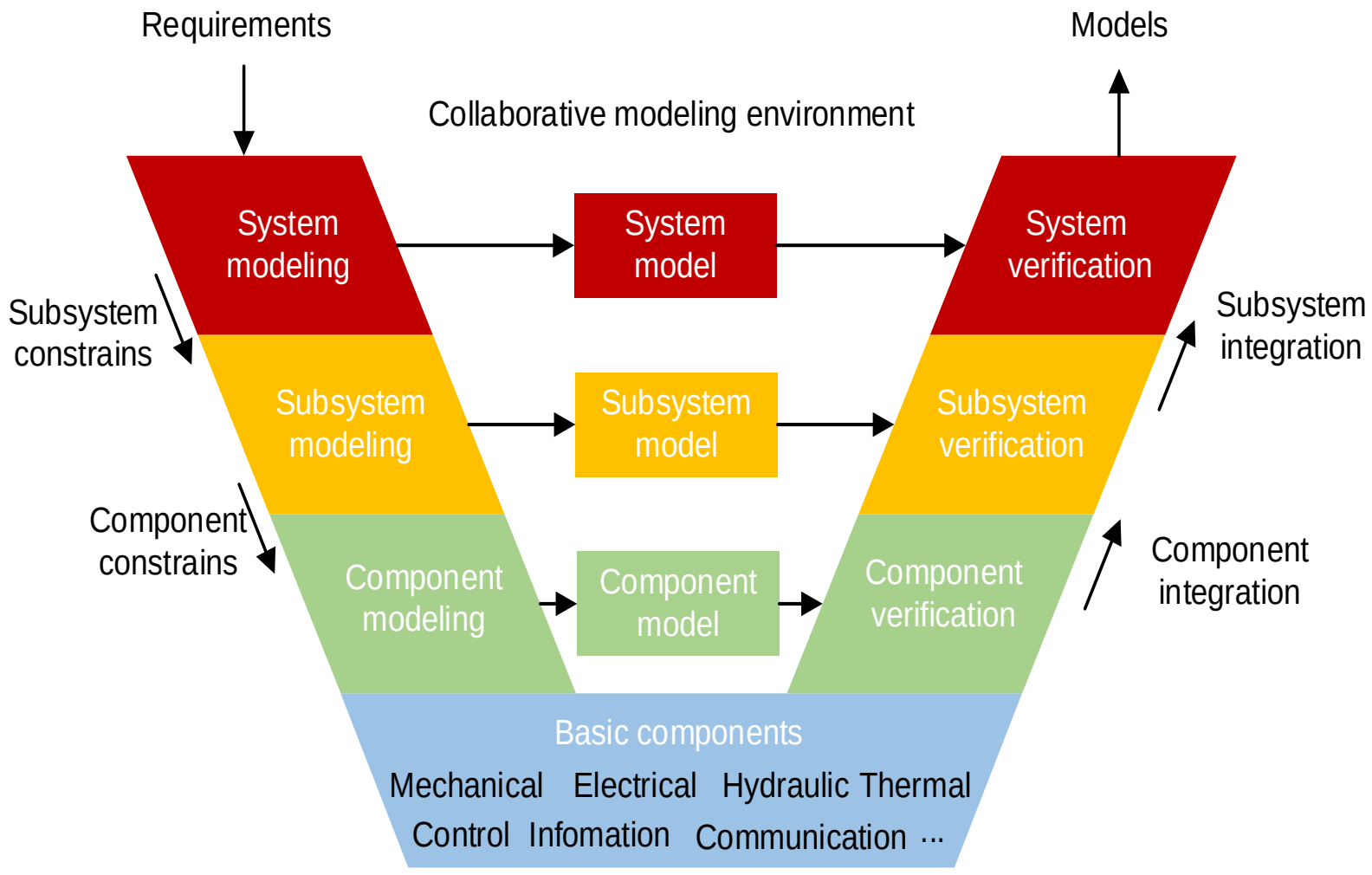

**Figure 1.** V-shaped process of system modeling.

A typical problem in EoM modeling tools is the state inconsistency in the simulation, which occurs when too many or too few equations are specified in the model [7,12]. An EoM with defects is called singular, which means that the underlying equation system has no unique and deterministic solution. The non-singularity of an EoM is the preliminary of system representation and simulation.

Structural analysis can verify the non-singularity of an EoM in the static analysis stage. Theoretically, the non-singularity of an EoM can be guaranteed by the numerical non-singularity of the underlying equation system. However, verifying the numerical non-singularity is very expensive, even as expensive as solving the equations when it comes to the algebraic equation models [7,12]. During the modeling, the singularity of a model should be fed back as quickly as possible. Therefore, in the static analysis stage, the non-singularity of structural analysis is assumed to be a sufficient condition for implying that the equation system has a unique and deterministic solution [12,13]. The structural analysis of EoMs is equivalent to analyzing the structure of the underlying equation systems. It depends on the correlation between the variables and equations, regardless of the numerical values of the variables.

The structural analysis of equation systems has been an important research area since the 1960s. In 1962, Steward reviewed the related works on equation system analysis and proposed basic concepts and methods to verify the solvability of an equation system with computers [14]. Subsequently, studies on structural analysis algorithms [15,16] and applications [17] have been conducted. These early works focused on partitioning large systems into small ones and testing if each of them was solvable [15,16]. The main principle to achieve this test was to permute the incidence matrix to a block lower triangular (BLT) matrix to make all diagonal elements non-zero. Based on the BLT form, the equations can be solved efficiently and sequentially with a forward substitution process [12]. The non-zero traversal of the incidence matrix is also considered a procedure that assigns each variable to a unique equation such that the variable appears in this equation [7,18]. If pairing variables and equations is impossible, then the equation system is structurally singular. The assignment method is always performed based on a bipartite graph representation. Equivalent to the matrix traversing method, this graph-represented method can use sparsity better to achieve improved performance in the sequential computation environment. However, these methods are only applicable to the structural analysis of algebraic EoMs expressing the static characteristics of the systems.

There are also works on the structural analysis for differential-algebraic equation (DAE) models that denote the dynamic characteristics of systems. Mattsson applied the assignment method for algebraic equations to DAEs without distinguishing between the appearances of a variable $x_i$ and its derivatives $\dot{x}_i$, $\ddot{x}_i$... [18]. This method is efficient but limited to catching singular models early, because a model can still be singular despite satisfying the assignment relation. The structural analysis of DAE models must consider the variable index and the initial conditions. Pantelides proposed a criterion for determining whether a subset of the equations should be differentiated to provide further constraints to the initial conditions [19]. His method is implemented as a graph-based algorithm to find consistent initial values for a DAE system. Unger derived the index reduction algorithm proposed by Gear [20] and presented a symbolical structural analysis algorithm based on the structural differentiation matrices [21]. The structural properties of DAEs, such as the solvability, dynamic degree of freedom and consistent initial condition, can be obtained by analyzing the structural differentiation matrices. Pryce proposed another matrix traversing approach to determine the highest order of derivatives to each equation and the highest indices of each variable based on the signature matrix [22]. This method converts the structural index problem into a maximum weight assignment problem to seek the highest-value traversal in the signature matrix. It is equivalent to the algorithm by Pantelides for the index-1 DAEs. The last 20 years have seen several extension works on developing the signature matrix-based structural analysis methods [23–27] and related tools [28]. However, these approaches only find, but cannot diagnose, the ill-posed model, because they will terminate their execution if a structural deficiency is encountered.

The diagnosis of structural singularity can be realized by Dulmage and Mendelsohn's (DM) decomposition algorithm [29,30]. Bunus realized DM decomposition with a graph-represented algorithm to find singular equations in a flat equation system [12]. Ding proposed a method to locate structural singularities in hierarchical Modelica models [31]. Their diagnosis method does not distinguish a variable and its derivatives in DAE systems. Soares realized a detailed diagnosis of DAEs by extending the graph-based algorithm by Pantelides [7]. This diagnosis obtains the information about the index and dynamic degree of freedom by augmenting the DAE system and finding a maximum matching in the bipartite graph of the augmented system. This method obtains a similar result with the signature matrix method presented in [22]. The DM decomposition algorithm can be applied to the augmented bipartite graph to diagnose the singularity source.

Although extensive works have been performed toward structural analysis, the computation of existing methods and tools is prohibitive when facing large-scale equations in the models of complex systems. For example, the variables and equations in a plane model can be millions or tens of millions in size. These approaches perform structural analysis based on the overall equation system obtained by flattening the hierarchical EoMs, thereby imposing challenges for analyzing the structural singularity of equations on such a scale. Some technical attempts, such as modeling and simulating different subsystems separately, verifying the DAE as nonlinear-algebraic equation (NLAE) systems [12,31] or decomposing the equations set into parts to analyze them separately [12] have been noted to address this challenge. However, the resultant defects such as the local optimal, low accuracy and additional computations from the decomposition are non-negligible for practical implementation.

In practical engineering, the EoMs are always modular and have a hierarchical structure. The components in a model are coupled with a few variables and equations. It is the natural sparse decomposition of an EoM. The structural analysis of complex EoMs can be carried out based on the natural hierarchical structure to avoid processing all the flattened equations at once. Based on this idea, this paper explores the relationship between the structural singularities of an EoM and its components and proposes a hierarchical structural analysis method. The proposed method can be adaptively applied to EoMs of different equation types. The hierarchical structural analysis of NLAE models that express static characteristics and DAE models that express dynamic characteristics are implemented as application cases. The main algorithms and the proof of the equivalence between the proposed method and existing methods based on flattened equations are presented. The efficiency of the proposed method is examined by application comparisons with the existing methods based on the flattened model. The time complexity analysis shows that the hierarchical structural analysis has better performance than the existing methods. Compared with existing structural analysis methods, the following distinguishing features should be noted:

1. Rather than performing the structural analysis based on the flattened equation model, the proposed method analyzes a hierarchical EoM based on a dummy model constructed by parts of each component.
2. The hierarchical analysis can be performed from the bottom up, layer by layer in the hierarchical model structure. This reduces the scale of equations in each step and enables the structural analysis of extremely complex EoMs.
3. The proposed method is more effective for hierarchical EoMs in practical engineering. It can be adaptively applied to NLAE models and DAE models.

The remainder of this paper is organized as follows. Section 2 provides a hierarchical abstraction of EoMs and introduces the basic concepts in the graph-represented structural analysis. Section 3 analyzes the relationship between the structural singularities between the model and its components and proposes the hierarchical structural analysis method. Section 4 applies the hierarchical structural analysis method to NLAE and DAE models. The main algorithms for analyzing the NLAE and DAE models are presented. The equivalence between the result of the proposed method and the existing methods is proven mathematically and verified with application examples. Section 5 discusses the time complexity of the hierarchical structural analysis method. The result is compared with the time complexity of the existing methods to prove its efficiencies. The advantages and disadvantages of the proposed method are discussed. Section 6 concludes this paper and gives possible directions for future research.

## 2. Preliminary

In this section, we abstract the hierarchical EoMs in different forms to describe them in a united fashion. Some basic concepts in the graph-represented structural analysis are also recalled. Table 1 gives the list of symbols throughout this paper.

Table 1. The list of symbols

| | Symbol | Meaning |
|---|---|---|
| 1 | $m$ | An equation-oriented model. |
| 2 | $\overline{m}$ | The flattened model of $m$. |
| 3 | $\widehat{m}$ | The dummy model of $m$. |
| 4 | $m_i^k$ | A $k$th-level component whose index is $i$. |
| 5 | $A$ | The set of variables defined in a model. It inherits the symbol marks from the model. |
| 6 | $R$ | The set of equations defined in a model. It inherits the symbol marks from the model. |
| 7 | $S$ | The set of components in a model. It inherits the symbol marks from the model. |
| 8 | $G$ | The bipartite graph of an equation-oriented model $m$. |
| 9 | $G^*$ | The bipartite graph of the augmented underlying ordinary differential equations of a DAE model. |
| 10 | $G^o$ | The over-constrained part of $G$. It inherits the symbol marks from the bipartite graph. |
| 11 | $G^u$ | The under-constrained part of $G$. It inherits the symbol marks from the bipartite graph. |
| 12 | $G^w$ | The well-constrained part of $G$. It inherits the symbol marks from the bipartite graph. |
| 13 | $A^o$ | The set of variables in the over-constrained part. It inherits the symbol marks from the graph. |
| 14 | $A^u$ | The set of variables in the under-constrained part. It inherits the symbol marks from the graph. |
| 15 | $A^w$ | The set of variables in the well-constrained part. It inherits the symbol marks from the graph. |
| 16 | $R^o$ | The set of equations in the over-constrained part. It inherits the symbol marks from the graph. |
| 17 | $R^u$ | The set of equations in the under-constrained part. It inherits the symbol marks from the graph. |
| 18 | $R^w$ | The set of equations in the well-constrained part. It inherits the symbol marks from the graph. |
| 19 | $M$ | A matching of a bipartite graph. |
| 20 | $M_{max}$ | A maximum matching of a bipartite graph. |

*2.1. Abstraction of Hierarchical Equation-Oriented Models*

An EoM, such as the Modelica model or the Simulink model, always includes a set of variables, a set of components that represent the subsystems and a set of equations that represent the relations between these variables and components. It can be abstracted as a triple $m = (A, S, R)$, where the following are true:

- $A$ is a finite set of variables that represents the states;
- $S$ is a finite set of components, also named submodels, that represent subsystems at a specific abstraction level;
- $R$ is a finite set of equations that represent the relation between the variables in $A$ and the variables in each component $m_i = (A_i, S_i, R_i) \in S$. Note that the variable in $A_i$ may appear in the equations in $R$.

An EoM without components is called a primary model. A component $m^1 = (A^1, S^1, R^1) \in S$ is called a first-level component. If a sequence $m^i = (A^i, S^i, R^i) \in S^{i-1}$ for $i = 1 \dots n$ exists, then the model $m^n = (A^n, S^n, R^n) \in S^{n-1}$ is called an $n$th-level component of $m$. Correspondingly, if all $n$th-level components of a model $m$ are primary models, then the model $m$ is called an $n$-level model. The variables and components in each level are associated with equations, thereby forming a hierarchical structure.

The hierarchical EoMs are flattened into a set of equations over a set of variables to perform the structural analysis [12]. Consider an $n$-level model $m_0 = (A, S, R)$. Define $A_0^* = A$ and $S_0^* = S$. For indices $i \in \{1 \dots n\}$, the sequence $\{A_i^*\}$ of variables in each level and the sequence $\{S_i^*\}$ of the components in each level can be defined such that

$$\begin{aligned} A_i^* &= \cup\{A_m | m = (A_m, S_m, R_m) \in S_{i-1}^*\}, \\ S_i^* &= \cup\{S_m | m = (A_m, S_m, R_m) \in S_{i-1}^*\}. \end{aligned} \tag{1}$$

Let $\bar{A} = \cup\{A_i^* | i \in \{0, \dots, n\}\}$. For the finite levels of $m_0$, $\bar{A}$ is the set that consists of all variables in $m_0$. Similarly, we can define $R_0^* = R$ and a sequence $\{R_i^*\}$ such that for $i \in \{1, \dots, n\}$

$$R_i^* = \cup\{R_m | m = (A_m, S_m, R_m) \in S_i^*\}. \tag{2}$$

Let $\bar{R} = \cup\{R_i^* | i \in \{0, \ldots, n\}\}$. Then, $\bar{R}$ is the exact unique set of equations in $m_0$. Therefore, an arbitrary EoM $m$ can be flattened into a primary model $\bar{m} = (\bar{A}, \emptyset, \bar{R})$ without components by induction of layers in the hierarchical structure.

In existing research, the structural analysis of a hierarchical EoM $m$ is performed based on the flattened model $\bar{m}$, which essentially is a set of equations. More details of this hierarchical abstraction of EoMs can be found in [2].

### 2.2. Concepts in Graph-Represented Structural Analysis

The term "structural singularity" in this paper follows the definition in [12,19]. Structural nonsingularity is a necessary condition for the uniqueness and existence of the solution. The "structural" portion means that the property is independent of the numerical values of the coefficients and variables in the equation. It has different meanings according to the type of equations in the model. For algebraic equation models, the structural singularity is equivalent such that the incidence matrix has no nonzero traversal. For ordinary differential equation (ODE) or DAE models, this means that the incidence matrix of the underlying ODEs formed by the index reduction process has no nonzero traversal. The formal definition of structural singularity for different type equation models will be present in the following sections.

The existing methods transform a hierarchical EoM into a flat equation system to perform the graph-represented or matrix-represented structural analysis [7,12,19,31]. The graph-represented structural analysis methods always have better performance in the sequential computation environment because of their better usage of the sparsity of practical EoMs. They represent the variables and equations in the model as a bipartite graph and analyze the graph to verify the existence and uniqueness of the solution. Here, some necessary concepts in the graph-represented structural analysis are introduced.

**Definition 1**: A bipartite graph $G$ of an equation model consists of a node set of all variables and equations in the model and an edge set $E$ indicating the presence of a variable in each equation.

If the model is a primary model without components, its bipartite graph is $G = (A \cup R, E)$. Otherwise, the bipartite graph is $G = (\bar{A} \cup \bar{R}, E)$, where $\bar{A}$ and $\bar{R}$ are from the flattened model $\bar{m} = (\bar{A}, \emptyset, \bar{R})$. Since the flattened form of a primary model is itself, the bipartite graph of an equation model can be generally denoted as $G = (\bar{A} \cup \bar{R}, E)$.

**Definition 2**: A matching $M$ of a bipartite graph is a subset of edges without common nodes.

**Definition 3**: A matching $M_{max}$ with maximum cardinality is called a maximum matching of the bipartite graph.

**Definition 4**: The nodes not covered by a matching $M$ are called exposed nodes of $M$.

**Definition 5**: If a matching covers all nodes of the variables and equations, then it is called a perfect matching.

Obviously, a perfect matching is a maximum matching. The exposed node set of a perfect matching is empty. Suppose a bipartite graph $G = (\bar{A} \cup \bar{R}, E)$ and a matching $M$ of $G$.

**Definition 6**: A path in $G$ is called an augment path of $M$ if it satisfies that (1) it has no repeated nodes; (2) its endpoints are exposed nodes of $M$; and (3) its edges are alternatively in $E \backslash M$ and in $M$.

Suppose that $M_{max}$ is a maximum matching of $G$. Then, the feasible path can be defined as follows.

**Definition 7**: A path in $G$ is called a feasible path if it satisfies that (1) it has no repeated nodes; (2) one of its endpoints is in $M_{max}$ and the other is an exposed node of $M_{max}$; and (3) its edges are alternatively in $E \backslash M_{max}$ and in $M_{max}$.

The graph-represented structural analysis needs to find a perfect matching or a maximum matching in the bipartite graph. A model with perfect matching is considered structurally non-singular [7]. Maximum and perfect matching can be found by classic augment path-searching algorithms [32–34]. These algorithms search augment paths from arbitrary matching (may be empty) until no new augment path can be observed. Notably, different maximum matchings can be found in a bipartite graph. Based on a maximum matching or perfect matching, new maximum matchings or perfect matchings can be found by altering the matching edges in a feasible path [35,36].

A model is considered structurally singular if its bipartite graph has no perfect matching. The source of singularity needs to be analyzed and reported to users. The canonical decomposition algorithm by Dulmage and Mendelsohn (*DM decomposition*) is used as one further step to reveal the source of singularity [12,29,31]. Based on a maximum matching $M_{max}$, the DM decomposition algorithm canonically decomposes a bipartite graph into three distinct parts: the over-constrained part $G^o$, the under-constrained part $G^u$ and the well-constrained part $G^w$. The sets of variables and equations in each part are denoted as $A^o$, $A^u$, $A^w$, $R^o$, $R^u$ and $R^w$.

Interestingly, a bipartite graph can admit different maximum matchings, but the final decomposition is unique (i.e., it does not depend on the choice of the maximum matching) [12,37]. Essentially, the DM decomposition considers all possible maximum matchings admitted by the bipartite graph. The over-constrained part $G^o$ is determined by the equations not covered by any possible maximum matchings. Similarly, the under-constrained part $G^u$ is determined by the variables not covered by any possible maximum matchings. The possible exposed variables and equations are searched by enumerating all possible maximum matchings. Although this task is apparently complex, it is indeed a very simple task. In graph theory jargon, this task is just a matter of finding feasible paths that begin at the uncovered nodes [35]:

$$\begin{cases} e1: f_1(v1) = 0 \\ e2: f_2(v1, v2) = 0 \\ e3: f_3(v2) = 0 \\ e4: f_4(v2, v3, v4) = 0 \\ e5: f_5(v3, v4) = 0 \\ e6: f_6(v4, v5, v6) = 0 \\ e7: f_7(v5, v6, v7) = 0 \end{cases} \tag{3}$$

An example equation system (Equation (3)) and the corresponding bipartite graph are presented in Figure 2 to illustrate these concepts. Equation (3) is considered a primary equation model $m = (A, \emptyset, R)$, where $A = \{v1, v2, v3, v4, v5, v6, v7\}$ is the variable set and $R = \{e1, e2, e3, e4, e5, e6, e7\}$ is the equation set. In the bipartite graph, the nodes in the upper and lower parts represent equations and variables, respectively. Each edge represents the presentation of a variable in an equation. The bold edges represent a maximum matching whose exposed nodes are $\{e1, v7\}$. Two feasible paths can be observed in the graph. One is $(v7, e7, v6, e6, v5)$ from the exposed variable node $v7$, and the other is $(e1, v1, e2, v2, e3)$ from exposed equation node $e1$. The equations in Equation (3) are decomposed into three parts, namely $R^o = \{e1, e2, e3\}$, $R^w = \{e4, e5\}$ and $R^u = \{e6, e7\}$, by applying the DM decomposition. The variables in each part are $A^o = \{v1, v2\}$, $A^w = \{v3, v4\}$ and $A^u = \{v5, v6, v7\}$, respectively. The nodes in different parts are painted in different colors.

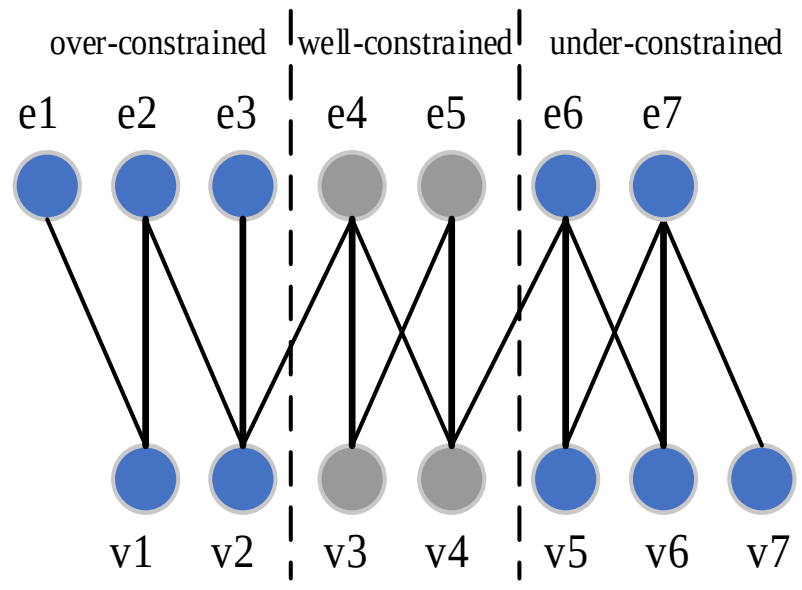

**Figure 2.** Decomposition of the bipartite graph representation of the equations in Equation (3).

## 3. Hierarchical Structural Analysis

The general process of existing structural analysis methods for EoMs is shown as the dashed arrow lines in Figure 3. The models in the hierarchical structure are flattened into an overall equation system to be analyzed and solved. During the structural analysis in a symbolic engine, the large-scale equations are decomposed into strong-connection components to reduce the computation complexity. However, the sparsity from the hierarchical structure is ignored in this procedure. This section will study the relationship between the singularities of a model and its components and propose a hierarchical method for structural analysis. As the solid arrow lines in Figure 3 show, the hierarchical structural analysis method can analyze hierarchical EoMs without flattening them.

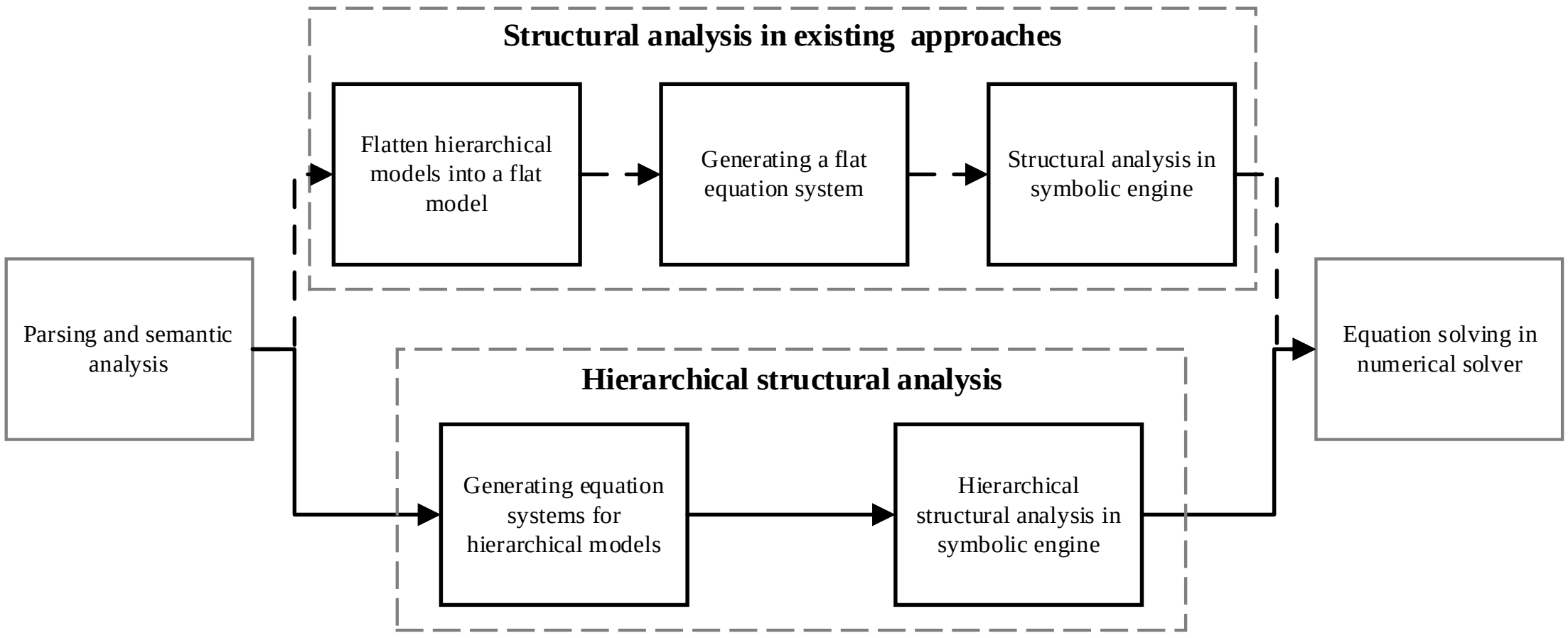

**Figure 3.** Comparison between the stages of the hierarchical structural analysis method and existing structural analysis methods.

The singularity of a model can be sourced into a subset of equations in it. The key to hierarchical structure analysis is to find an equation subset, which may be the singularity source, in each component. Thus, the structural singularity of a model can be revealed by analyzing these "picked" equations. The equation subset should be unique for a specified component so it is the intrinsic property of the component and will not be changed no matter how the equation subset is obtained or where the component is used. More importantly, the union of the equation subsets should include all possible singularity sources of the model.

DM decomposition can be used to diagnose the possible singularity sources of the components. As introduced in Section 2.2, it decomposes the bipartite graph of an equation system into three distinct parts: the over-constrained part $G^o$, the under-constrained part $G^u$ and the well-constrained part $G^w$. The decomposing result is unique for each bipartite graph [12,30,37]. The singularity of a model is caused by the equations in $G^o$ and $G^u$ [7,12,31]. Therefore, the equation subset for hierarchical structural analysis can be picked up from a component by DM decomposition.

The components in a hierarchical EoM are always predefined. They should contain only the under-constrained part and the well-constrained part. A component with an over-constrained part is not allowed because the redundant equations will make the model singular. Therefore, the components are assumed to be well-constrained or under-constrained during the structural analysis of a hierarchical EoM.

In the structural analysis of a model $m = (A, S, R)$, the under-constrained part and the well-constrained part of a component $m_i = (A_i, S_i, R_i) \in S$ have different effects. The under-constrained part needs constraints from equations in $m$, whereas the well-constrained part provides constraints to the common variables in $m_i$ and $m$. The structural analysis of $m$ can regard the variables in the well-constrained part of $m_i$ as known variables or independent functions that respect the time and further analyze the equations in $R$ and the under-constrained part of $m_i$. The structural singularity of the model $m$ can be revealed by structural analysis of the union set of equations in $m$ and the under-constrained part of each component $m_i$.

In summary, hierarchical structural analysis of an EoM can be performed as illustrated by the process in Figure 4. Each component in a model is represented as a bipartite graph and decomposed into distinct parts. The equations in the model and the under-constrained parts of the components make up a dummy model to reveal the structural singularity of the model. The equivalence between the structural singularities of the dummy model and the original model will be mathematically analyzed in a graph-theoretical context in the next section. The analysis process can be performed on hierarchical EoMs layer by layer to verify the structural singularity of very complex models.

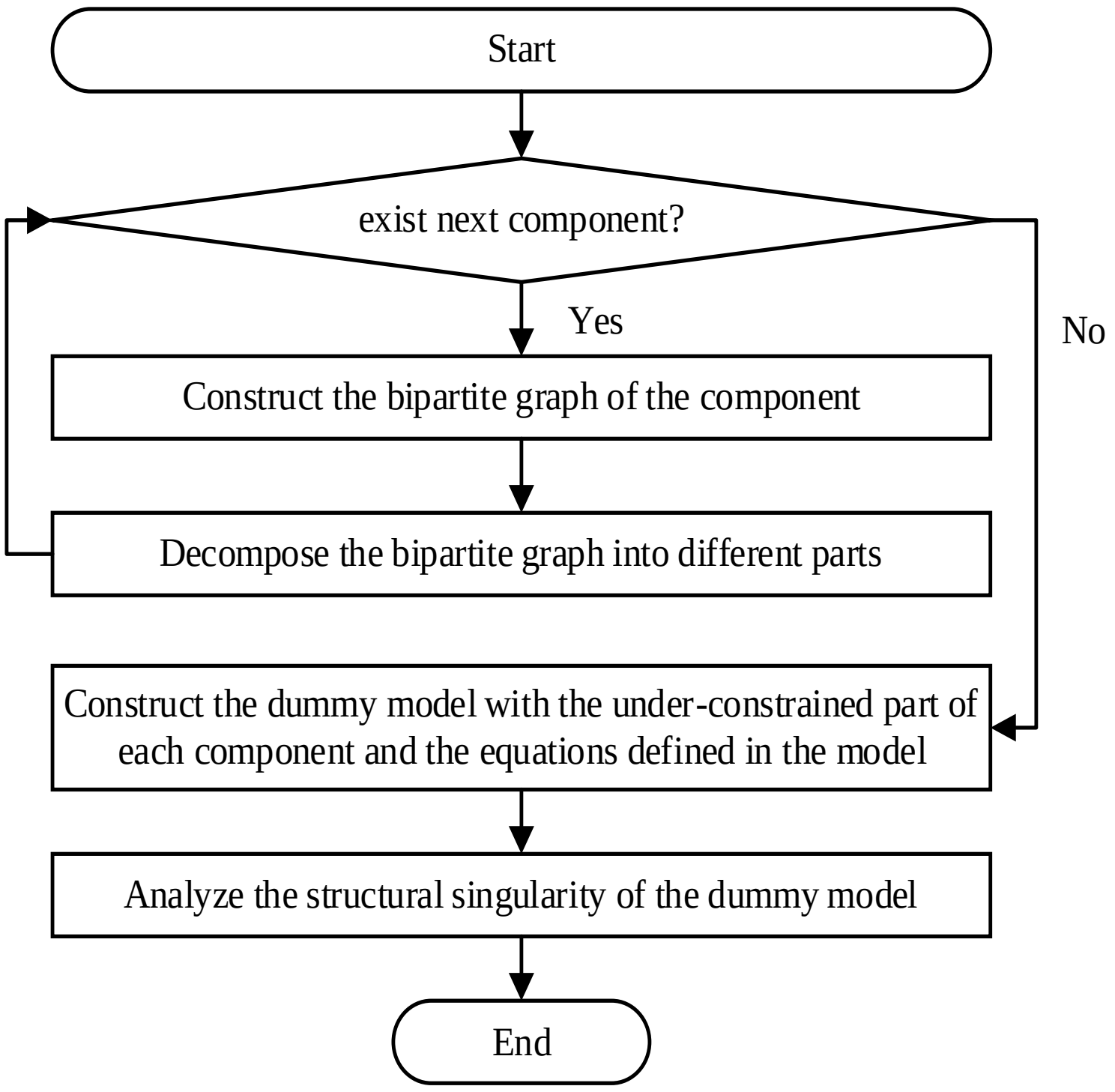

**Figure 4.** Process of hierarchical structural analysis.

## 4. Hierarchical Structural Analysis of NLAE and DAE Models

In practical implementation, the detail of hierarchical structural analysis is different depending on the types of equations in the model. The structural singularity of NLAE models is obtained by analyzing the incidence matrix. The structural analysis of DAE models is different from that of NLAE models, because the indices of the variables should be considered. This section will implement hierarchical structural analysis for NLAE and DAE models and apply the algorithms to the practical structural analysis of NLAE and DAE models. The simpler NLAE models are considered first to explain our idea clearer.

### *4.1. NLAE Models*

As stated in Section 2, a hierarchical NLAE model can be flattened into a primary model $\bar{m} = (\bar{A}, \emptyset, \bar{R})$, which is essentially an NLAE equation system:

$$\mathbf{F}(\boldsymbol{x}) = \mathbf{0}. \tag{4}$$

The structural singularity of $\bar{m}$ is equivalent to the structural singularity of Equation (4). The structural singularity of Equation (4) can be analyzed by finding a nonzero traversal in the incidence matrix or finding a perfect matching in the bipartite graph [15,16].

In the graph-represented structural analysis approaches, Equation (4) is represented as a bipartite graph $G = (\bar{A} \cup \bar{R}, E)$, where $E$ is the set of edges representing the presence of a variable in each equation and $\bar{A}$ and $\bar{R}$ are the variables and equations in Equation (4), respectively. Therefore, the structural singularity of an NLAE model can be defined as follows.

**Definition 8:** An NLAE model is called structurally singular if its bipartite graph has no perfect matching.

#### 4.1.1. Decomposition of Components

Hierarchical structural analysis performs DM decomposition on each component to obtain their under-constrained parts. DM decomposition can be realized by finding feasible paths in the directed bipartite graph based on a maximum matching. The feasible paths divide the bipartite graph into three distinct parts: the over-constrained part where the nodes and edges are sourced from exposed equation nodes, the under-constrained part where the nodes and edges target exposed variable nodes, and the well-constrained part where the nodes and edges are outside the feasible paths. More details on graph-represented DM decomposition are presented in [12,35,36].

Algorithm 1 is our variant implementation of DM decomposition. The difference between algorithm 1 and other implementations, such as Bunus' implementation in [12] and Ding's implementation in [31], is that algorithm 1 only considers the feasible paths starting at exposed variable nodes. It assumes that each component is consistent and has no over-constrained part. Moreover, algorithm 1 adopts a similar skill to the Hopcroft–Karp Matching (*HKMatching*) algorithm [32] to speed up searching for feasible paths. In each loop, it finds multiple disjoint length-2 feasible paths from the exposed variables. Subsequently, the alternating variable nodes on these feasible paths are appended into the under-constrained variable set and treated as the exposed variables in the next loop.

The pseudocode of decomposing an NLAE component is presented as follows.

**Algorithm 1**: Decomposition of an NLAE component.

Input the bipartite graph $G(\bar{A} \cup \bar{R}, E)$ of the component.

Output variables $A^u$ and equations $R^u$ in the under-constrained part.

1: let $M_{max}$ be a maximum matching, set $M_{max}$ = HKMatching$(G)$;
2: let $A^e$ be a queue of exposed variables, $A^e = \bar{A} - \{a \mid \exists r \in R((a,r) \in M_{max})\}$;
3: if $A^e = \emptyset$, return $\emptyset$, $\emptyset$;
4: let $A^u$ be the set of under-constrained variables, $A^u = A^e$;
5: let $G^d$ be the corresponding directed graph of $G$, $G^d$ = direct$(G, M_{max})$;
6: while $A^e \neq \emptyset$:
7:    let $a$ be an element in $A^e$ and remove it from $A^e$;
8:    let $P$ be the length-2 feasible paths from $a$, $P$ = feasiblePath$(G^d, a)$;
9:    for each $p = (a, r, a') \in P$:
10:       if $a' \in A^u$, continue;
11:       append $a'$ into $A^e$
12:       append $a'$ into $A^u$, $A^u = A^u \cup \{a'\}$;
13: let $R^u$ be the equation set in the under-constrained part, $R^u = \{r \mid \exists a \in A^u((a,r) \in M_{max})\}$
14: return $A^u$, $R^u$

Our remarks on algorithm 1 are as follows:

1. This algorithm assumes that the component has no over-constrained part.
2. In line 2, the function HKmatching$(G)$ is the maximum matching algorithm by Hopcroft and Karp [32]. It can stably find a maximum matching of the bipartite graph $G$ in $O\left(\sqrt{|\bar{A}| + |\bar{R}|} * |E|\right)$. This step can be realized with other maximum matching algorithms, such as the algorithms in [33,34,38] with time complexity $O((|\bar{A}| + |\bar{R}|) * |E|)$.
3. In line 3, the notation $\emptyset$ indicates an empty set. If the bipartite graph has no exposed variables, the sets of variables and equations in the under-constrained part are empty.
4. In line 5, the function direct$(G, M_{max})$ directs the edges to form a directed bipartite graph $G^d = (\bar{A} \cup \bar{R}, E^d)$, where $E^d = \{(r \in \bar{R}, a \in \bar{A}) \mid (a,r) \in M_{max}\} \cup \{(a \in \bar{A}, r \in \bar{R}) \mid (a,r) \in E - M_{max}\}$. The directions of the edges are used to state the searching process clearer. They are optional in the practical implementation.
5. In line 8, the function feasiblePath$(G^d, a)$ begins from each exposed variable $a \in A^e$ and finds length-2 feasible paths in $G^d$. Searching of each feasible path needs to access two nodes. The time complexity of this function is $2 * c$, where $c$ is the average number of edges for each node.
6. Lines 6–12 find all under-constrained variables $A^u$ from the exposed variables. Each under-constrained variable is appended in the queue and popped from the queue only once. Therefore, the time complexity is within $(2 + 2 * c + 2) * |A^u|$.

Algorithm 1 decomposes an NLAE component into the under-constrained part and well-constrained part. Take the system of equations *e*4, *e*5, *e*6, and *e*7 in Equation (3) as an example. As shown in Figure 5, the bipartite graph of the equations is decomposed into two parts by algorithm 1. The bold edges in the bipartite graph represent a maximum matching. All edges are directed depending on whether they are in the maximum matching. In the directed bipartite graph, the variable node *v*7 is not covered by the matching. A feasible path (*v*7, *e*7, *v*6) can be found from *v*7. It indicates that *v*6 is an under-constrained variable. From *v*6, a feasible path (*v*6, *e*6, *v*5) can be found. Then, *v*5 is also an under-constrained variable. From *v*5, a length-2 path (*v*5, *e*6, *v*6) exists. It is not a feasible path, because *v*6 has been in the under-constrained part. Therefore, the under-constrained variable set can be denoted as $A^u = \{v5, v6, v7\}$. The under-constrained equation set $R^u$ is the set of equations matched to the under-constrained variables.

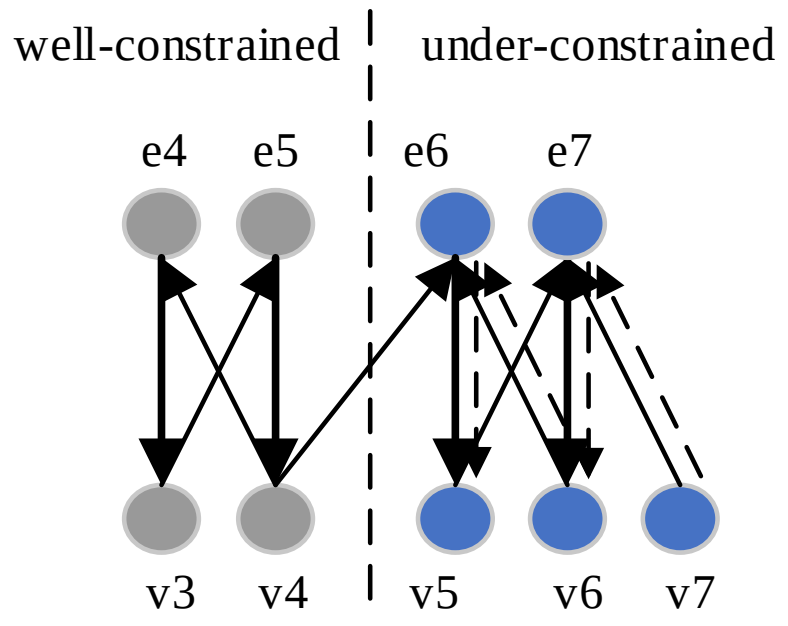

**Figure 5.** Feasible paths from exposed variable $v7$.

Under the assumption that the component has no over-constrained nodes, lemma 1, 2, and 3 hold.

**Lemma 1.** *If an equation is in the well-constrained part of a component, then its variables are also in the well-constrained part.*

**Proof.** Assume that a well-constrained equation $r \in R^w$ is matched to a well-constrained variable $a \in A^w$.

According to the assumption that the component has no over-constrained nodes, the variables in $r$ are well-constrained or under-constrained.

If an arbitrary variable $a'$ in $r$ is under-constrained, then a feasible path $(a', r, a)$ exists. The feasible path indicates that the equation $r$ and the variable $a$ are in the under-constrained part. It contradicts the assumption that $r$ and $a$ are well-constrained.

Therefore, an arbitrary variable in a well-constrained equation is well-constrained, and lemma 1 is proven.

**Lemma 2.** *If a well-constrained model is augmented with a set of equations, the over-constrained part of the equation set is a subgraph of the over-constrained part of the augmented model.*

**Proof.** Assume a well-constrained model $m = (A, S, R)$ and an equation system $R'$ with variables $A'$.

Denote the bipartite graph of $m$ as $G = (A \cup R, E)$. It has at least one perfect matching $M$. Each equation $r_i \in R$ is matched to a unique variable $a_j \in A$ and vice versa.

Denote the bipartite graph of $R'$ as $G' = (A' \cup R', E')$ and the bipartite graph of the augmented model as $G^* = (A \cup A' \cup R \cup R', E \cup E')$. For a maximum matching $M'$ of $G'$, the union $M \cup M'$ is a maximum matching of $G^*$.

All nodes in $G$ are covered by the maximum matching $M \cup M'$. No new augment path exists in $G^*$. The over-constrained nodes in $G'$ are still over-constrained in $G^*$.

Therefore, the over-constrained part of $G'$ is a subgraph of the over-constrained part of $G^*$, and lemma 2 is proven.

**Lemma 3.** *If a well-constrained model is augmented with a set of equations, the under-constrained part of the equation set equals the under-constrained part of the augmented model.*

**Proof.** Assume a well-constrained model $m = (A, S, R)$ and an equation system $R'$ with variables $A'$.

Denote the bipartite graph of $m$ as $G = (A \cup R, E)$. $G$ has at least one perfect matching $M$. Each equation $r_i \in R$ is matched to a unique variable $a_j \in A$ and vice versa.

Denote the bipartite graph of $R'$ as $G' = (A' \cup R', E')$ and the bipartite graph of the augmented model as $G^* = (A \cup A' \cup R \cup R', E \cup E')$. For a maximum matching $M'$ of $G'$, the union $M \cup M'$ is a maximum matching of $G^*$.

Direct the edges in $G^*$ as $E^d = \{(r, a) | \{a, r\} \in M \cup M'\} \cup \{(a, r) \big| \{a, r\} \in (E \cup E' - (M \cup M'))\}$. All edges between $G$ and $G'$ start at the variables in $G$ and end at the equations in $G'$. For the under-constrained variables in $G'$, no feasible path enters $G$. Therefore, no new under-constrained node can be observed in $G^*$.

Moreover, no new augment path exists in $G^*$, because all nodes in $G$ are covered by the maximum matching $M \cup M'$. No new matching edge can be found. Therefore, the under-constrained nodes in $G'$ are still in the under-constrained part of $G^*$.

Therefore, the under-constrained part of $G^*$ equals the under-constrained part of $G'$, and lemma 3 is proven.

#### 4.1.2. Construction of the Dummy Model

For a hierarchical EoM $m = (A, S, R)$, the dummy model is constructed based on the decomposing result of each component. The equations in the dummy model are a subset of the equations in the flattened model $\bar{m}$. The structural analysis of the dummy model can reveal the structural singularity of the original model $m$.

**Definition 9**: The *dummy model* of an NLAE model $m = (A, S, R)$ is defined as $\hat{m} = \left(A \cup (\bigcup_{i \in S} A_i^u), \emptyset, R \cup (\bigcup_{i \in S} R_i^u)\right)$, where $A_i^u$ and $R_i^u$ are variables and equations in the under-constrained part of each component, respectively.

The pseudocode of constructing the dummy model for an NLAE model is presented as follows.

**Algorithm 2**: Construction of the dummy model $\hat{m}$.

Input a model $m = (A, S, R)$; output the dummy model $\hat{m}$.

```
1:   Let Â = A, R̂ = R;
2:   for each m_i = (A_i, S_i, R_i) ∈ S, do
3:     G_i = bipartiteGraph(m_i)
4:     A_i^u, R_i^u = decompose(G_i);
5:   let Â = Â ∪ A_i^u;
6:   let R̂ = R̂ ∪ R_i^u;
7:   let m̂ be the dummy model of m, m̂ = (Â, ∅, R̂);
```

Our remarks on algorithm 2 are as follows:

1. In line 3, the function constructs a bipartite graph for the component $m_i$.
2. In line 4, the function $\text{decompose}(m_i)$ is an implementation of algorithm 1. It decomposes a component $m_i$ and returns the variable set and the equation set in the under-constrained part.
3. If the component set is empty, the dummy model is equivalent to the flattened model.

**Theorem 1.** *For an NLAE model, the structural singularities of the dummy model and the flattened model are equivalent.*

**Proof.** Assume an NLAE model $m = (A, S, R)$. The corresponding flattened model is denoted as $\bar{m} = (\bar{A}, \emptyset, \bar{R})$, where $\bar{A}$ and $\bar{R}$ represent the union sets of the variables and equations in $m$ and its components. The dummy model of $m$ is denoted as $\hat{m} = (\hat{A}, \emptyset, \hat{R})$.

If $m$ is a primary model, the component set $S$ is empty. Obviously, $\hat{A} = A = \bar{A}$ and $\hat{R} = R = \bar{R}$. The dummy model and the flattened model contain the same variables and equations. Their structural singularities are equivalent.

If $m$ is a first-level model, each component $m_i = (A_i, \emptyset, R_i) \in S$ is a primary model. Algorithm 1 decomposes each component $m_i$ into the under-constrained part $G_i^u = (A_i^u \cup R_i^u, E_i^u)$ and the well-constrained part $G_i^w = (A_i^w \cup R_i^w, E_i^w)$.

According to definition 9 and the assumptions $A_i = A_i^u \cup A_i^w$ and $R_i = R_i^u \cup R_i^w$, the dummy model $\hat{m}$ satisfies the following:

$$\begin{aligned} \hat{A} &= A \cup (\textstyle\bigcup_{i \in S} A_i^u) = A \cup (\bigcup_{i \in S} A_i) - (\bigcup_{i \in S} A_i^w) = \bar{A} - (\bigcup_{i \in S} A_i^w), \\ \hat{R} &= R \cup (\textstyle\bigcup_{i \in S} R_i^u) = R \cup (\bigcup_{i \in S} R_i) - (\bigcup_{i \in S} R_i^w) = \bar{R} - (\bigcup_{i \in S} R_i^w). \end{aligned} \tag{5}$$

For each component $m_i$, a well-constrained model $\mathrm{m_i^w} = (\mathrm{A_i^w}, \emptyset, \mathrm{R_i^w})$ can be built with the variables and equations in the well-constrained part. According to lemma 2, the over-constrained part of $\hat{m}$ is a subgraph of the over-constrained part of the model $m_i^{w*} = \left(A_i^w \cup \hat{A}, \emptyset, R_i^w \cup \hat{R}\right)$. Under the assumption that the component has no over-constrained nodes, the over-constrained parts of $\hat{m}$ and $m_i^{w*}$ are equivalent. According to lemma 3, the under-constrained part of $\hat{m}$ equals the under-constrained part of $m_i^{w*}$. Therefore, the structural singularities of $\hat{m}$ and $m_i^{w*}$ are equivalent.

Inductively, the structural singularity *of* $\hat{m}$ equals that of the model $m_S^w = \left((\bigcup_{i \in S} A_i^w) \cup \hat{A}, \emptyset, (\bigcup_{i \in S} R_i^w) \cup \hat{R}\right)$. Because $\hat{A} \subseteq \bar{A}$, $\hat{R} \subseteq \bar{R}$, $A_i^w \subseteq \bar{A}$ and $R_i^w \subseteq \bar{A}$ and the contents Equation (5), $m_S^w = (\bar{A}, \emptyset, \bar{R}) = \bar{m}$.

Therefore, the structural singularity of the first-level model $m$ is equivalent to that of its dummy model $\hat{m}$.

If $m$ is an $n$-level $(n \geq 2)$ model, it can be converted into a first-level model $m' = (A, \bar{S}, R)$ satisfying $\forall \bar{m}_i \in \bar{S}\left(\bar{m}_i = (\bar{A}_i, \emptyset, \bar{R}_i)\right)$. The structural singularities of $m'$ and $m$ are equivalent because they have the same variables and equations.

The dummy model of $m'$ satisfies $\hat{m}' = \left(A \cup (\bigcup_{i\in S} \bar{A}_i^u), \emptyset, R \cup (\bigcup_{i\in S} \bar{R}_i^u)\right)$. The equivalence of the structural singularities of the first-level model $m'$ and its dummy model has been proven. Therefore, the structural singularities of $m$ and $\hat{m}'$ are equivalent.

According to definition 9, the dummy model of $m$ is $\hat{m} = (\hat{A}, \emptyset, \hat{R}) = \left(A \cup (\bigcup_{i\in S} A_i^u), \emptyset, R \cup (\bigcup_{i\in S} R_i^u)\right)$. For each component $m_i$, the under-constrained part of the dummy model is equal to the that of the flattened form $\bar{m}_i$, satisfying $A_i^u = \bar{A}_i^u$ and $R_i^u = \bar{R}_i^u$, and therefore

$$\hat{A} = A \cup (\bigcup_{i\in S} A_i^u) = A \cup (\bigcup_{i\in S} \bar{A}_i^u),$$

$$\hat{R} = R \cup (\bigcup_{i\in S} R_i^u) = R \cup (\bigcup_{i\in S} \bar{R}_i^u).$$

The dummy models of $m$ and $m'$ are equal, and the structural singularities of $\hat{m}$ and $\hat{m}'$ are equivalent.

Therefore, the structural singularities of the $n$-level ($n \geq 2$) model $m$, the component-flattened model $m'$, the dummy model $\hat{m}'$ and the dummy model $\hat{m}$ are equivalent, and theorem 1 is proven.

Theorem 1 ensures the equivalence of the dummy model and the flattened model for structural singularity verification and diagnosis. Note that this is only established if it assumes that the components have no over-constrained nodes. The over-constrained part of the dummy model is a subgraph of the over-constrained part of the flattened model. They can be equivalent when ignoring the over-constrained nodes in the components. However, if desired, the over-constrained nodes in a component can be found by searching the feasible paths in the bipartite graph augmented by the assignment equations of the variables appearing in both the component and the over-constrained parts of the dummy model.

Note that the dummy model can be built based on the dummy models of the components. Denote the dummy model of the component $m_i = (A_i, S_i, R_i) \in S$ as $\hat{m}_i = (\hat{A}_i, \emptyset, \hat{R}_i)$. According to lemma 3, the under-constrained part of $m_i$ and $\hat{m}_i$ are equal, satisfying $\hat{A}_i^u = A_i^u$ and $\hat{R}_i^u = R_i^u$. Therefore, the dummy model $\hat{m}$ satisfies

$$\hat{A} = A \cup (\bigcup_{i\in S} A_i^u) = A \cup \left(\bigcup_{i\in S} \hat{A}_i^u\right),$$

$$\hat{R} = R \cup (\bigcup_{i\in S} R_i^u) = R \cup \left(\bigcup_{i\in S} \hat{R}_i^u\right).$$

$\hat{A}_i^u$ and $\hat{R}_i^u$ can be obtained by hierarchical structural analysis of the components. This means that the analysis results of the components can be used to build the dummy model of $m$. This ensures that the hierarchical structural analysis can be performed iteratively layer by layer in the hierarchical structure of EoMs.

#### 4.1.3. Hierarchical Structural Analysis Case of NLAE Models

Theorem 1 indicates that structural analysis of the dummy models can reveal the structural singularity of the hierarchical EoMs. The structural singularity of NLAE models can be obtained by performing the structural analysis method, based on maximum matching searching or DM decomposition, on the dummy model.

Take the system in Figure 6 as an example to illustrate the hierarchical structural analysis of the NLAE models. This system consists of a heat-generating circuit and a shell dissipating heat into the environment. The variables and equations in the models can be found in the dataset [39]. By applying algorithm 1 on the circuit component and the shell component, the graphs in Figure 7a,b are obtained. In each bipartite graph, the bold edges represent a maximum matching. The gray nodes and the blue nodes represent the well-constrained parts and the under-constrained parts of the components, respectively. The dummy model can be constructed by performing algorithm 2 on each component. Figure 7c shows the result of applying the DM decomposition on the dummy model. All nodes in the graph are well-constrained, which indicates that the system model is well-posed. As a comparison, Figure 7d gives the result of applying the DM decomposition algorithm on the flattened model, where all nodes are also well-constrained. The comparison of Figure 7c,d shows that the hierarchical structural analysis method is effective and can obtain an equivalent singularity result. The resulting graphs imply that the proposed method can reduce the node scale in structural analysis of NLAE models.

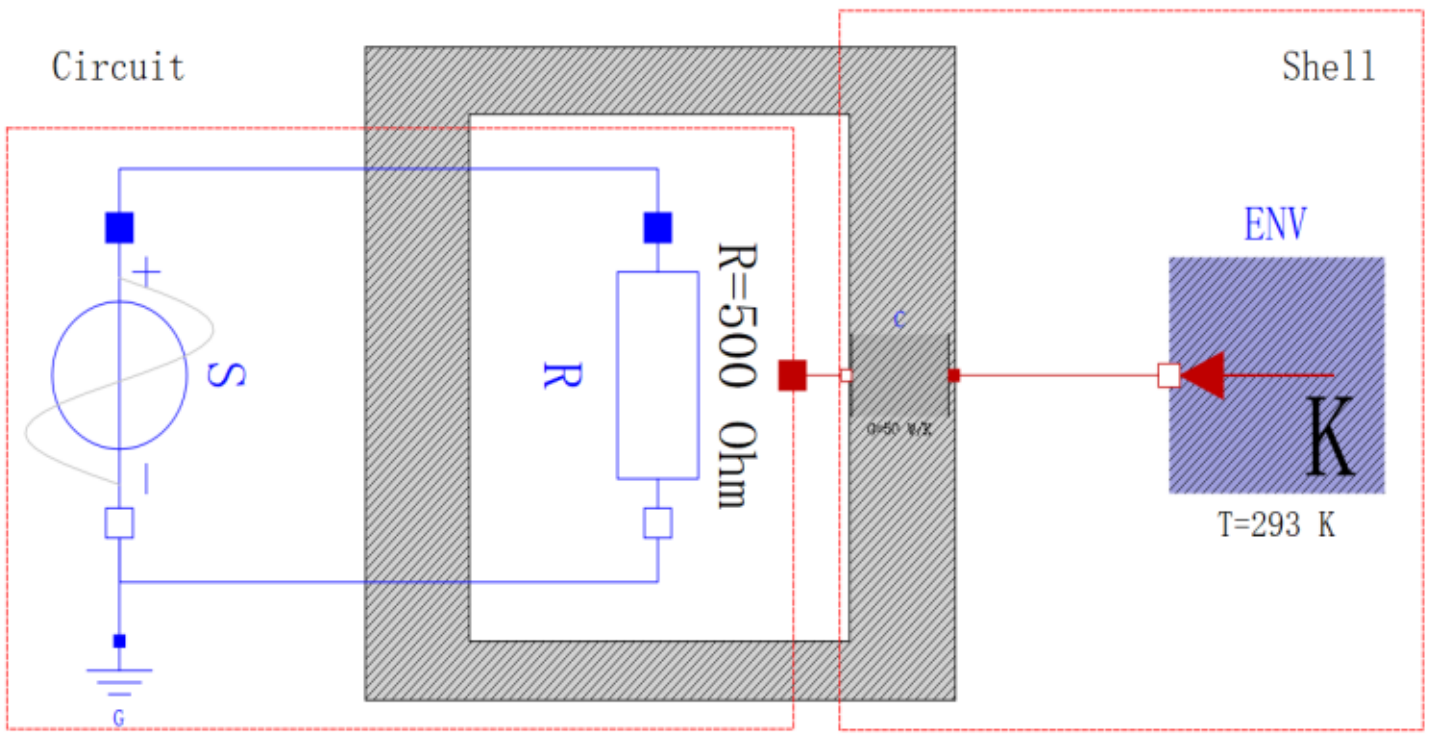

**Figure 6.** Example system to illustrate the structural analysis of NLAE models.

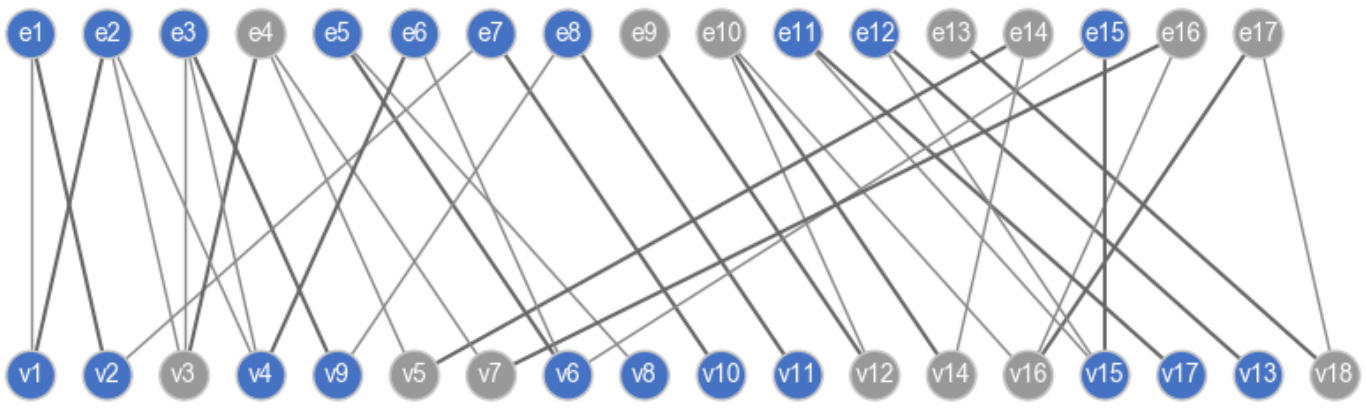

(**a**) Decomposition of the circuit model.

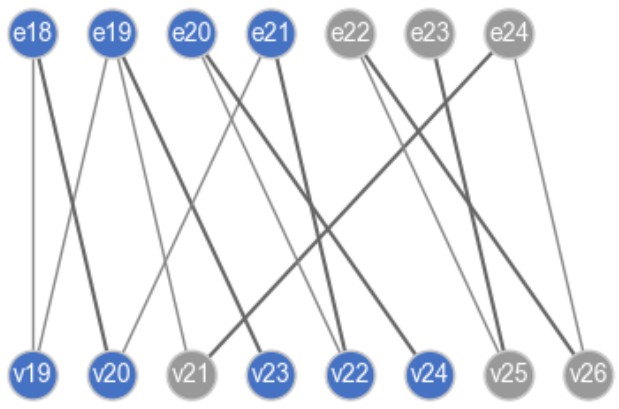

(**b**) Decomposition of the shell model.

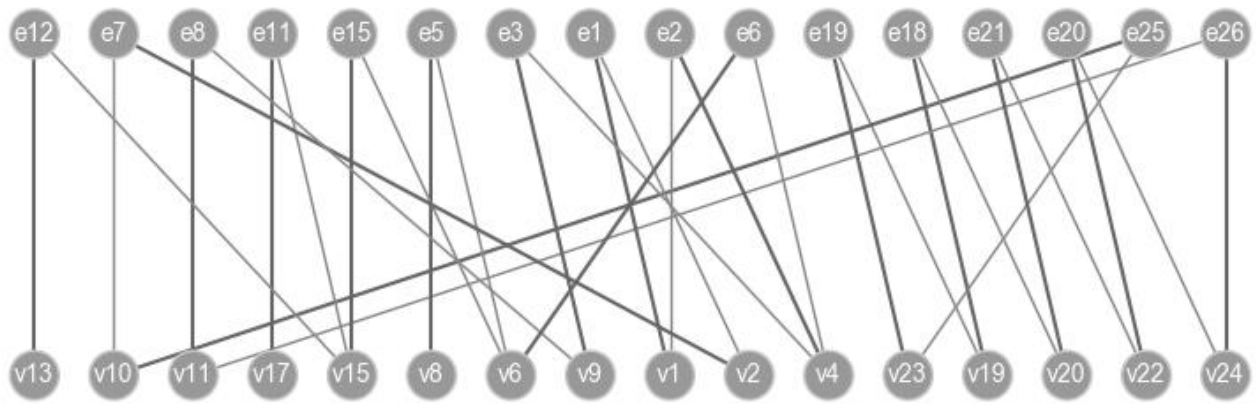

(**c**) Structural analysis result of the dummy model.

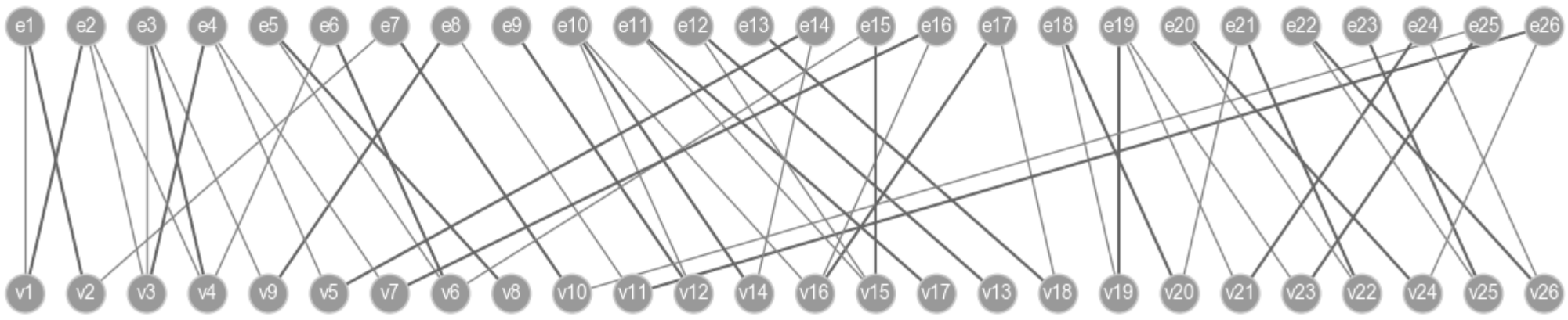

(**d**) Structural analysis result of the flattened model.

**Figure 7.** Hierarchical structural analysis of the NLAE model in Figure 6.

### *4.2. DAE Models*

A hierarchical DAE-oriented model is essentially a DAE system. Assuming that the equations are infinitely differentiable, a DAE system can be equivalently augmented into an implicit underlying ODE (UODE) system in Equation (6) by an index reduction process [13,20]. Note that the equations in Equation (6) only contain the variables and their first-order derivatives:

$$\mathbf{F}(x, \dot{x}, t) = \mathbf{0} \tag{6}$$

Equation (6) is finally transformed into an ordinary differential equation (ODE) system $\dot{\boldsymbol{x}} = \mathbf{F}_1(\boldsymbol{x}, t)$ to be solved by the numerical methods. The solvability of the UODE system demands the consistency of the differentiated variables $\dot{\boldsymbol{x}}$. The UODE augmented with the equations reduced in the index reduction process is used to solve the initial value problem. The graph-represented methods are always used to efficiently verify the consistency of $\dot{\boldsymbol{x}}$ and the initial values [7].

In graph-represented approaches, the consistency of the variables is verified by a procedure that assigns each equation to a unique variable. The variables that need initialization are determined by the exposed variables in the bipartite graph of the augmented UODE (AUODE) system. The AUODE can be considered an NLAE by replacing the derivatives with independent algebraic variables, similar to the dummy derivative method by Mattsson [13]. A consistent AUODE is always under-constrained and needs constraints from the initial conditions. Therefore, the structural singularity of a DAE model can be defined in a graph-theoretical context as follows.

**Definition 10:** A DAE model is called structurally singular if the bipartite graph of its AUODE system has an over-constrained part.

The structural analysis of a DAE model aims to find the redundant equations of the model and the variables that require initialization. This section will implement hierarchical structural analysis on the DAE models to build a dummy model. More preliminary knowledge about DAEs, such as index reduction, consistent initial values, UODE and AUODE, can be found in [7,12,13,20,31].

#### 4.2.1. Decomposition of Components

Different from NLAE models, DAE models are structurally analyzed based on the AUODE obtained by an index reduction process. Hierarchical structural analysis decomposes the bipartite graph of the resulting AUODE of the components into different parts. The graph-represented algorithm by Soares and Secchi (SSMatching) [7] is adopted to augment the DAE system into the AUODE and find a maximum matching in the augmented bipartite graph.

The SSMatching algorithm extends the graph-represented algorithm by Pantelides. It selects equations that require differentials by prioritizing the index consistency and differentiates them in each loop. The algorithm is iteratively executed until the indices of the variables are consistent or a deficiency is found. The criterion for selecting equations and the termination condition ensures that the resulting equation system is minimal. The resulting augmented bipartite graph is equivalent to the bipartite graph of the AUODE of the DAE system.

In hierarchical structural analysis, the components of a hierarchical DAE model are augmented and decomposed to obtain the under-constrained part. The pseudocode of the DAE component decomposition algorithm is listed as follows.

**Algorithm 3:** Decomposition of a DAE component.

Input a bipartite graph $G(\bar{A} \cup \bar{R}, E)$.

Output the variables $A^u$ and equations $R^u$ in the under-constrained part.

let $G^* = (\bar{A^*} \cup \overline{R^*}, E^*)$ be the bipartite graph of the augmented DAEs; let $M^*_{max}$ be a maximum matching, $G^*, M^*_{max} = \text{SSMatching}(G)$.
let $A_e$ be a queue of exposed variables, $A^e = A^* - \{a | (\exists r \in R^*)(a, r) \in M^*_{max}\}$;
if $A_e$ is empty, return $\emptyset$, $\emptyset$;
set $G^{*d} = \text{direct}(G^*, M^*_{max})$;
let $A^u$ be the set of under-constrained variables, $A^u = A^e$;
while $A^e \neq \emptyset$:
  let $a$ be an element in $A^e$ and remove $a$ from $A^e$;
  let $P$ be the length-2 feasible paths from $a$, $P = \text{feasiblePaths}(G^{*d}, a)$;
  for each $p = (a, r, a') \in P$:
    if $a' \in A^u$, continue;
    append $a'$ into $A^e$
    append $a'$ into $A^u$, $A^u = A^u \cup \{a'\}$;
let $R^u$ be the equation set in the under-constrained part, $R^u = \{r | \exists a \in A^u((a, r) \in M^*_{max})\}$
return $A^u$, $R^u$

Our remarks on algorithm 3 are as follows:

1. In line 1, the function SSMatching($G$) augments the bipartite graph $G$ and finds a maximum matching in the augmented bipartite graph $G^* = (A^* \cup R^*, E^*)$. The SSMatching algorithm needs at most $O(n^3)$ operations to find a maximum matching for a DAE system.
2. The difference between algorithm 3 and algorithm 1 is that decomposing a DAE model is performed on the augmented bipartite graph $G^*$ rather than the bipartite graph $G$ of the original equation system.

Take the DAE model in Equation (7), which is also an example in [7,40], as an example. The bipartite graph of this model is shown in Figure 8a. After applying the SSMatching algorithm, the equations $e2$, $e3$ and $e4$ are differentiated, and a maximum matching (bold edges) is found. The augmented bipartite graph is shown in Figure 8b. In Figure 8c, algorithm 3 decomposes the augmented graph into two parts, where the nodes are colored in gray and blue. The nodes in the set $\{V, V', e2, e2'\}$ are in the well-constrained part, and the remaining nodes are in the under-constrained part:

$$\begin{aligned} &e1: \dot{U} + P * \dot{V} = u_1(t) \\ &e2: V = u_2(t) \\ &e3: P * V = n * R * T \\ &e4: U - U_0 = n * c_v * T \end{aligned} \tag{7}$$

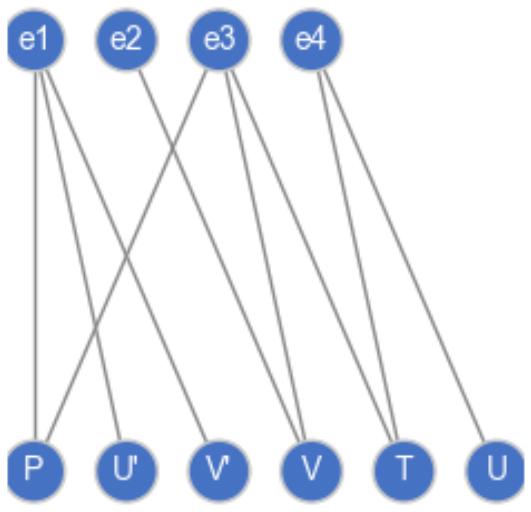

(**a**) Original bipartite graph.

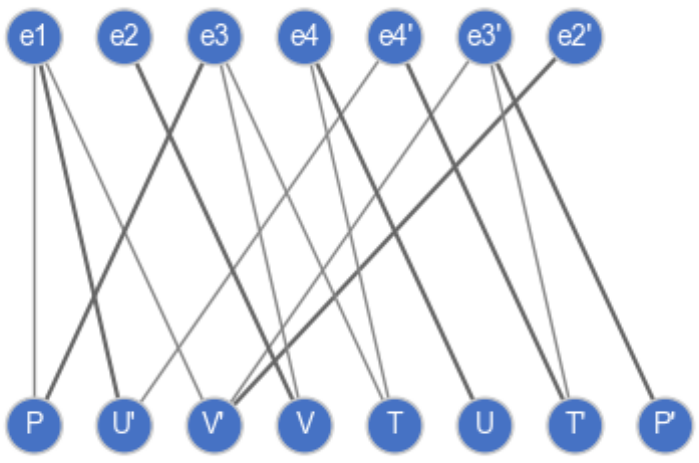

(**b**) Augmented bipartite graph.

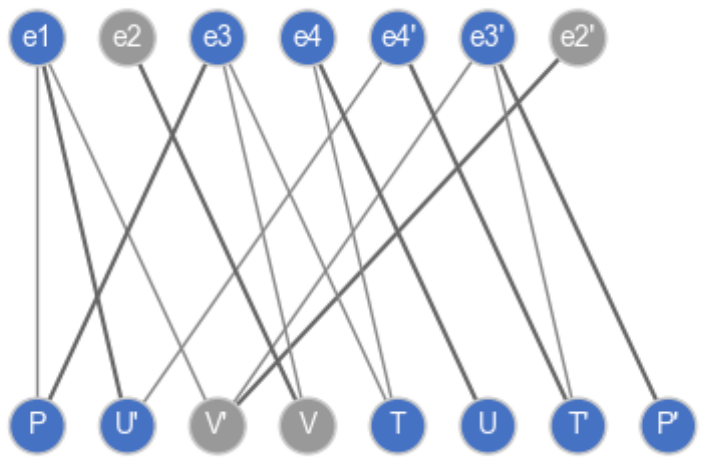

(**c**) Decomposition of the model.

**Figure 8.** Decomposition of DAEs in the model of Equation (7).

### 4.2.2. Construction of the Dummy Model

In hierarchical DAE models, the derivatives of the variables in the components may appear in the equations defined in the model. During construction of the dummy model for a DAE model, these derivatives must be handled properly. In a DAE model, if a variable is the derivative of a variable in the well-constrained part of a component, it should be ignored in the dummy model like the well-constrained variables. The reason for this is that if a variable in a DAE system is well-constrained, then all of its derivatives are also well-constrained under the assumption that the component has no over-constrained nodes.

**Lemma 4.** *If a variable is well-constrained in a DAE system, an arbitrary derivative of the variable is well-constrained in a system augmented by differentiating equations, and the under-constrained parts of the original system and the augmented system are equivalent.*

**Proof.** Assume that the bipartite graph of the AUODE system is $G = (A, \emptyset, R)$ and the decomposing results of $G$ are $G^u = (A^u \cup R^u, E^u)$ and $G^w = (A^w \cup R^w, E^w)$.

In $G^w$, each equation $e_i \in R^w$ is matched to a unique variable $a_j \in A^w$. Differentiating each equation $e_i \in R^w$ produces a graph $G'$ with a set of equations $\{e_i'\}$ and a set of variables $\{\dot{v}_j\}$, and $\dot{v}_j$ is a variable in $e_i'$. There exists a matching $\{(e_i', \dot{v}_j)\}$ in the bipartite graph $G'$.

According to lemma 1, all variables in $e_i$ are in $A^w$. Therefore, the variables in $\{e_i'\}$ are the elements of $\{\dot{v}_j\}$. Each variable in $G'$ is matched to a unique equation. Let $G^* = G \cup G'$. The variables and equations in $G'$ are in the well-constrained part of the augment bipartite graph $G^*$.

Therefore, for a variable $v \in A^w$, its derivative $\dot{v} \in \{\dot{v}_j\}$ is well-constrained in the augmented DAE system represented by $G^*$.

Since the variables in $A^w$ and the equations in $R^w$ are in $G^w$, differentiating them will not change the variables and equations in $G^u$. Therefore, $G^u = G^{*u}$.

Inductively, the derivative $\ddot{v}$ is well-constrained in the augmented DAE system represented by $G^{**} = G^* \cup G^{*'}$. For the arbitrary index $d$, the derivative $v^d$ is well-constrained in the augmented DAE system represented by $G^{(d+1)}$, and the under-constrained parts of $G^*$ and $G^{d+1}$ are equivalent.

Therefore, lemma 4 is proven.

Based on the decomposing result of the components, the dummy model of a DAE model can be built under the following definition.

**Definition 11**: The dummy model of a DAE model $m = (A, S, R)$ is defined as $\hat{m} = \left(A \cup (\cup_{i \in S} A_i^u) \ominus (\cup_{i \in S} A_i^w), \emptyset, R \cup (\cup_{i \in S} R_i^u)\right)$, where the operator $X \ominus Y$ means removing all variables in $Y$ and their derivatives from $X$, and $A_i^u$ and $R_i^u$ are the variables and equations in the under-constrained part of each component, respectively.

The pseudocode of the dummy model's construction is presented as follows.

| **Algorithm 4:** Construction of dummy model $\hat{m}$ | |
|---|---|
| Input a model $m = (A, S, R)$; output the dummy model $\hat{m}$. | |
| 1: | Let $\hat{A} = A$, $\hat{R} = R$; |
| 2: | for each $m_i = (A_i, S_i, R_i) \in S$, do |
| 3: | $G_i = \text{bipartiteGraph}(m_i)$ |
| 4: | $A_i^u, R_i^u = \text{decompose}(G_i)$; |
| 5: | let $\hat{A} = \hat{A} \cup A_i^u \ominus A_i^w$; |
| 6: | let $\hat{R} = \hat{R} \cup R_i^u$; |
| 7: | let $\hat{m}$ be the dummy model of $m$, set $\hat{m} = (\hat{A}, \emptyset, \hat{R})$; |

Note that the derivatives of the variables in a DAE component $m_i$ may appear in the equations $R$. These derivatives should be treated as variables in the component $m_i$ rather than in the model $m$. The well-constrained variables in a component and their derivatives are removed from the dummy model in line 5. The handling of variable derivatives is the difference between algorithms 2 and 4.

**Theorem 2.** *The structural singularities of a DAE model and its dummy model are equivalent.*

**Proof.** Assume a DAE model $m = (A, S, R)$. The flattened form of $m$ is $\bar{m} = (\bar{A}, \emptyset, \bar{R})$. According to definition 10, the structural singularity of $m$ is equivalent to the structural singularity of the augmented model $\bar{m}^* = (\bar{A}^*, \emptyset, \bar{R}^*)$.

According to definition 11, the dummy model $\hat{m} = (\hat{A}, \emptyset, \hat{R})$ of $m$ is constructed with the decomposition result of the components $m_i = (A_i, S_i, R_i) \in S$.

If $m$ is a primary model, the variables and equations in $\bar{m}$ equal the variables and equations in $\hat{m}$. Their structural singularities are equivalent.

If $m$ is a first-level model, each component $m_i = (A_i, \emptyset, R_i) \in S$ is a primary model. The decomposed result of a component $m_i$ satisfies $A_i^u = A_i^{*u}$, $R_i^u = R_i^{*u}$, $A_i^w = A_i^{*w}$ and $R_i^w = R_i^{*w}$, where the superscript $*$ represents that they are in the bipartite graph of the AUODE of $m_i$.

If no new derivatives of the variables in $A_i^w$ appear in $A$, $\hat{A} = A \cup (\bigcup_{i\in S} A_i^u)$, then $\overline{m}^*$ can be considered the sum of the AUODE of the components and the AUODE of the equations in $R$. The structural singularity of the dummy model is equivalent to the structural singularity of the model $\widehat{m}^* = \left(A^* \cup (\bigcup_{i\in S} A_i^{*u}), \emptyset, R^* \cup (\bigcup_{i\in S} R_i^{*u})\right)$. When considering $\overline{m}^*$ as a first-level NLAE model, $\widehat{m}^*$ is the dummy model of the NLAE model. According to theorem 1, the under-constrained part and the over-constrained part of $\overline{m}^*$ and the dummy model constructed with the decomposition results are equivalent.

If the derivatives of the variables in $A_i^w$ appear in $A$, the variables in the dummy model satisfy the following:

$$\hat{A} = A \ominus \left(\bigcup_{i\in S} A_i^{*w}\right) \cup \left(\bigcup_{i\in S} A_i^{*u}\right)$$

Assume a well-constrained variable $v \in A_i^w$ and a derivative $v^d$ in an equation in $R$. According to lemma 4, the derivative $v^d$ is the well-constrained part of an AUODE of an augmented model $m_i^{**}$. The under-constrained parts of $m_i^{**}$ and $m_i^*$ satisfy $A_i^{**u} = A_i^{*u}$ and $R_i^{**u} = R_i^{*u}$. Their well-constrained parts satisfy $A_i^{*w} \subseteq A_i^{**w}$ and $R_i^{*w} \subseteq R_i^{**w}$. Then, the model $m$ can be considered a variant model $m'$, whose components are the augmented model $m_i^{**}$ and have no variable derivatives appearing in $A$. Therefore, the structural singularities of $m$ and $\widehat{m}'$ are equivalent.

The dummy model $\widehat{m}'$ satisfies

$$\hat{A}' = A \cup (\textstyle\bigcup_{i\in S} A_i^{**u}) = A \ominus (\bigcup_{i\in S} A_i^{*w}) \cup (\bigcup_{i\in S} A_i^{**u}) = A \ominus (\bigcup_{i\in S} A_i^{*w}) \cup (\bigcup_{i\in S} A_i^{*u}) = \hat{A},$$

$$\hat{R}' = R \cup (\textstyle\bigcup_{i\in S} R_i^{**u}) = R \cup (\bigcup_{i\in S} R_i^{*u}) = \hat{R},$$

Therefore, the dummy models $\widehat{m}$ and $\widehat{m'}$ are equal. The structural singularities of the dummy model $\widehat{m}$ and the flattened model $\overline{m}$ are equivalent.

The $n$-level model can be considered a first-level model by flattening its components. Therefore, the structural singularity of the DAE model $m$ and the dummy model $\widehat{m}$ are equivalent, and theorem 2 is proven.

### 4.2.3. Hierarchical Structural Analysis Case of the DAE Models

The structural singularity of the DAE models can be analyzed based on their dummy models. Take the system in Figure 9 as an example. This system consists of a heater component and a driver component connected with a vessel filled with gas. Its model is a hierarchical DAE model. The variables and equations in the model are presented in the dataset [39].

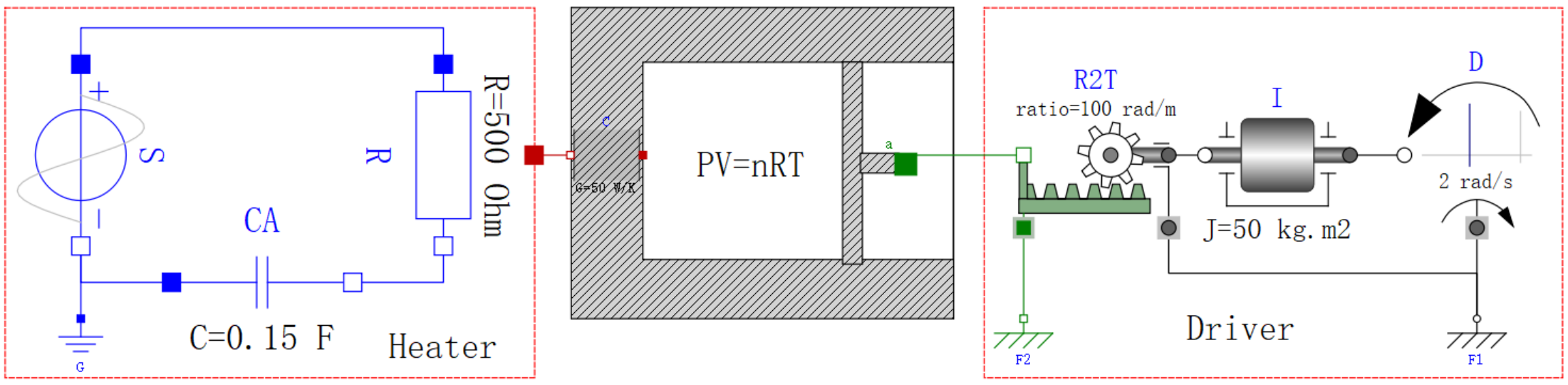

**Figure 9.** Example system to illustrate the hierarchical structural analysis of DAE models.

Table 2. Variables in different parts of the dummy model and the flattened model.

| | **Variables in the Under-Constrained Part $A^u$** | **Variables in the Well-Constrained Part $A^w$** |
|---|---|---|
| Flattened model $\overline{m}$ | [v18, v19, v61′, v52′, v13, v10, v11, v17, v18″, v15, v1′, v21′, v19′, v8′, v60″, v35′, v29′, v11′, v65, v64, v63, v61, v60, v10″, v6′, v11″, v18′, v55′, v9′, v10′, v4″, v63′, v7′, v64″, v33′, v52″, v23′, v53″, v54′, v20″, v13′, v9″, v55″, v59, v64′, v47′, v15′, v51″, v31′, v4′, v47, v17′, v65′, | [v62″, v12, v16, v48′, v14, v46‴, v40′, v62′, v62, v49′, v48‴, v45″, v22′, v44″, v34′, v48″, v41′, v5″, v49″, v26′, v46′, v12′, v42″, v46″, v41″, v32′, v24′, v40″, v57′, v43″, v44′, v50″, v49‴, v44‴, v57″, v25′, v41, v40, v43, v42, v45, v44, v46, v49, v48, v22″, v45‴, v14′, v26, v45′, |
| | v56′, v39′, v54″, v23, v21, v20, v29, v2′, v59′, v56, v54, v55, v52, v53, v51, v58, v51′, v31, v33, v35, v37, v39, v58′, v3′, v3″, v59″, v37′, v53′, v60′, v1, v2, v3, v4, v6, v7, v8, v9, v20′, v2″, v7″, v1″] | v24, v28, v25″, v5′, v30′, v28‴, v38″, v57, v50, v30, v32, v34, v36, v22, v38, v38′, v32″, v27, v36″, v28″, v12″, v24″, v25, v42′, v27′, v50′, v26″, v16′, v5, v14″, v30″, v16″, v36′, v43′, v34″, v28′, v25‴] |
| Dummy model $\hat{m}$ | [v18, v19, v61′, v52′, v13, v10, v11, v17, v18″, v15, v1′, v21′, v19′, v8′, v15′, v35′, v29′, v11′, v65, v64, v63, v61, v60, v10″, v6′, v11″, v18′, v55′, v9′, v10′, v4″, v63′, v7′, v64″, v33′, v52″, v23′, v53″, v54′, v20″, v13′, v9″, v51′, v64′, v47′, v60″, v55″, v51″, v31′, v4′, v47, v39, v65′, v56′, v39′, v54″, v23, v21, v20, v29, v2′, v59′, v56, v54, v55, v52, v53, v51, v58, v59, v31, v33, v35, v37, v17′, v58′, v3′, v3″, v59″, v37′, v53′, v60′, v1, v2, v3, v4, v6, v7, v8, v9, v20′, v2″, v7″, v1″] | [v62″, v57, v62′, v62, v57″, v57′] |
| Differences between variables in each part of the models | [] | [v12, v16, v48′, v14, v46‴, v40′, v49′, v48‴, v45″, v22′, v44″, v34′, v48″, v41′, v5″, v49″, v26′, v46′, v12′, v42″, v46″, v41″, v32′, v24′, v40″, v43″, v44′, v50″, v49‴, v44‴, v25′, v41, v40, v43, v42, v45, v44, v46, v49, v48, v22″, v45‴, v14′, v26, v45′, v24, v28, v25″, v5′, v30′, v28‴, v38″, v50, v30, v32, v34, v36, v22, v38, v38′, v32″, v27, v36″, v28″, v12″, v24″, v25, v42′, v27′, v50′, v26″, v16′, v5, v14″, v30″, v16″, v36′, v43′, v34″, v28′, v25‴] |

Algorithm 3 decomposes the models of the heater and driver into distinct parts filled with colors, as shown in Figure 10a,b. The under-constrained parts of the components, combined with the equations in the top-level model, construct the dummy model of the top-level system. DM decomposition is performed on the dummy model to reveal the possible over-constrained equations and the variables that require initialization. The decomposing result is shown in Figure 10c. As a comparison, the structural analysis of the corresponding flattened model is shown in Figure 10d.

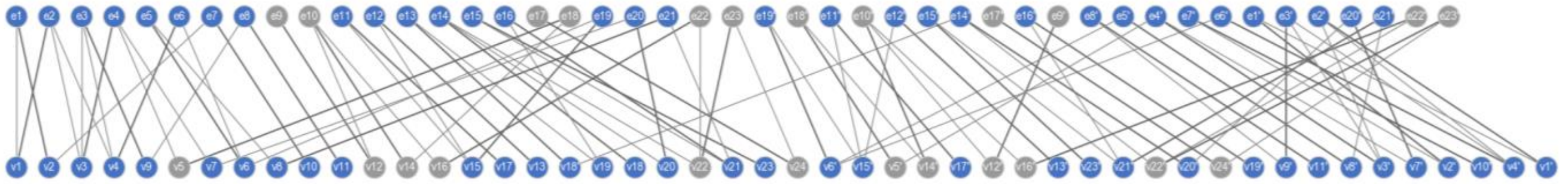

(**a**) Decomposition of the circuit model.

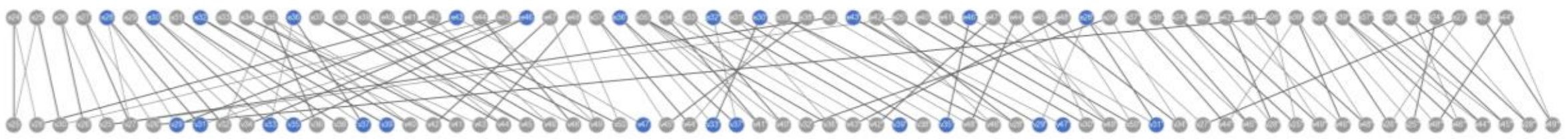

(**b**) Decomposition of the driver model.

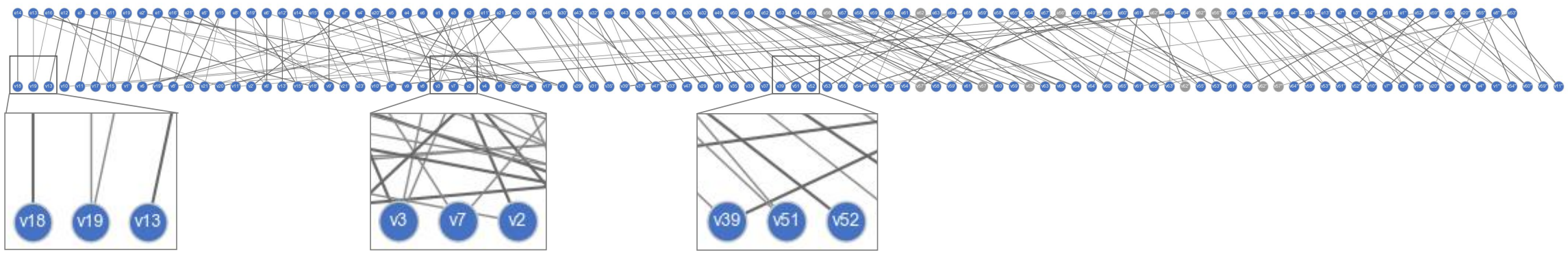

(**c**) Structural analysis result of the dummy model.

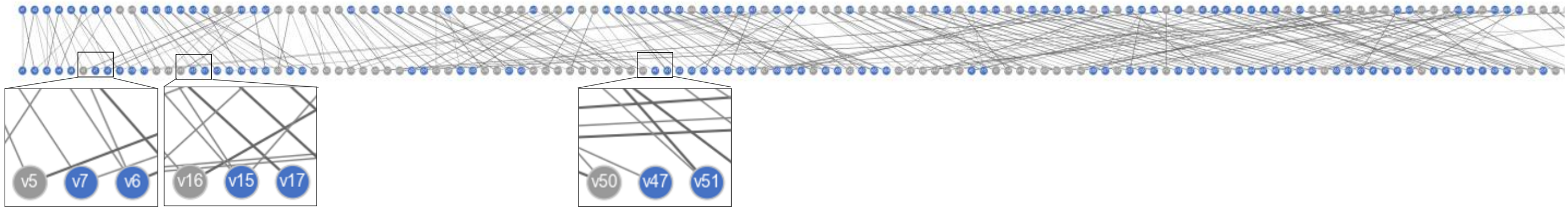

(**d**) Structural analysis result of the flattened model.

**Figure 10.** Structural analysis of the DAE model in Figure 9.

The variables in each part of the dummy model and the flattened model are listed in Table 2. The exposed variable set of the dummy model is {$v$19, $v$7, $v$51}, whereas the exposed variable set of the flattened model is {$v$15, $v$7, $v$47}. As shown in Figure 10d, $v$19 and $v$15 are connected by the feasible path ($v$15, $e$19, $v$6, $e$5, $v$8, $e$21, $v$21, $e$16, $v$19), and $v$47 and $v$51 are connected by the feasible path ($v$47, $e$43, $v$29, $e$28, $v$31, $e$46, $v$35, $e$30, $v$33, $e$36, $v$37, $e$32, $v$39, $e$63, $v$63, $e$57, $v$58, $e$54, $v$60, $e$59, $v$64, $e$60, $v$55, $e$50, $v$51). Therefore, the result of the hierarchical structural analysis is equivalent to that of the structural analysis based on the flattened model. We can select one from each pair of the alternating variables on the same feasible path as the variables that require initialization. For example, the variables {$v$7, $v$15, $v$51} can be chosen as the initial condition of the model. They represent the negative pin voltage of the resistor, the current through the source and the temperature difference at the ends of the vessel, respectively. When comparing Figure 10c,d, the nodes in the dummy model are much lower than those in the flattened model. The hierarchical structural analysis could achieve better performance than the existing structural analysis method.

## 5. Complexity Analysis and Discussion

The hierarchical structural analysis method includes three stages: decomposing the components, constructing the dummy model and performing structural analysis on the dummy model. The time complexity of each stage can be estimated as follows.

Decomposing an NLAE component takes three steps: (1) finding a maximum matching, (2) searching the feasible paths and determining the under-constrained variable set and (3) computing other parts by set operation. In step 1, the best sequential algorithm for finding a maximum matching in a bipartite graph is due to Hopcroft and Karp [32]. Their algorithm solves the maximum matching problem in $O(\sqrt{n} * m) = O(n^{5/2})$, where $n = |A \cup R|$ is the number of nodes and $m = |E|$ is the number of edges. Step 2 searches the under-constrained variables $A^u$. Its time cost is $u * (2 * c + 4)$, where $u$ is the number of nodes in the under-constrained part and $c$ is the average number of edges to each node. Step 3 computes the variables and equations in the well-constrained part and removes them and the related edges from the graph. The operations are in $u + (n - u) * (c + 1)$. Therefore, the total time cost of the component decomposition is $\sqrt{n} * m + n * (c + 1) + u * (c + 2)$. Considering that $m \leq n^2$ and $0 \leq u \leq n$, the time cost of the component decomposition is a function of $n$:

$$\begin{aligned} \mathrm{C}(n) &\leq n^{5/2} + n * (c + 1) + u * (c + 2) \\ &\leq n^{5/2} + (2 * c + 3) * n \end{aligned} \quad (8)$$

The dummy model is constructed based on the decomposing results of the components. Assuming a model with $k$ components, the node number in the dummy model is $\mathrm{n}_0 + \sum_1^k \mathrm{u}_\mathrm{i}$, where $n_0$ is the number of variables and equations in the model and $u_i$ is the number of nodes in the under-constrained part of each component. The time cost of constructing the dummy model is $c * (n_0 + \sum_1^k u_i)$, where $c$ is the average number of edges of a node.

The time cost of the last stage depends on the structural analysis algorithm adopted. Take the singularity diagnosis algorithm in [31] as an example. The number of nodes in the dummy model is $n_0 + \sum_1^k u_i$. Performing the algorithm on the dummy model requires $(n_0 + \sum_1^k u_i)^{5/2} + c * (n_0 + \sum_1^k u_i)$ operations at most.

Equation models in practical engineering are sparse, such that $c \ll n$. Therefore, the total time cost for the hierarchical structural analysis of an NLAE model is

$$\begin{aligned} C_{\text{total}} &= \sum_1^k \mathrm{C}(n_i) + c * \left(n_0 + \sum_1^k u_i\right) + \left(n_0 + \sum_1^k u_i\right)^{5/2} + c * \left(n_0 + \sum_1^k u_i\right) \\ &\leq \sum_1^k n_i^{5/2} + (2 * c + 3) * \sum_1^k n_i + 2 * c * \left(n_0 + \sum_1^k u_i\right) + \left(n_0 + \sum_1^k u_i\right)^{5/2} \\ &\leq \sum_1^k n_i^{5/2} + c_0 * \sum_1^k n_i + \left(n_0 + \sum_1^k u_i\right)^{5/2}. \end{aligned} \tag{9}$$

In practical system modeling, the components are always reused in different models. Their under-constrained and well-constrained parts can be decomposed previously. The pre-computed decomposing results will reduce the time cost of the hierarchical structural analysis to

$$C_{\text{reuse}} \leq \left(n_0 + \sum_1^k u_i\right)^{5/2} + c_0 * \sum_1^k u_i. \tag{10}$$

The existing structural analysis methods, such as the methods in [12,31], are based on flattened models. The time cost for flattening the hierarchical models is $O(|A \cup R|)$. Similar to the complexity analysis of the proposed method, the diagnosis algorithm in [31] is used here to analyze the structural singularity of a flattened model. The time complexity of the analysis is $O\left(|A \cup R|^{5/2}\right)$ [31]. Therefore, the best time cost for the existing structural singularity analysis is

$$C_{\text{flattened}} \leq \left(\sum_0^k n_i\right)^{5/2} + \sum_0^k n_i \tag{11}$$

Define $r = u/n$ as the ratio of under-constrained nodes and $c_0 = 6$ as the average number of edges to each node. According to Equations (9) and (10), we can plot the time complexities of the hierarchical structural analysis method in different situations by varying the variable number $n$, the component number $k$ and the under-constrained ratio r. In Figure 11, the results are compared with the time complexity of existing structural analysis methods at the same variable scale.

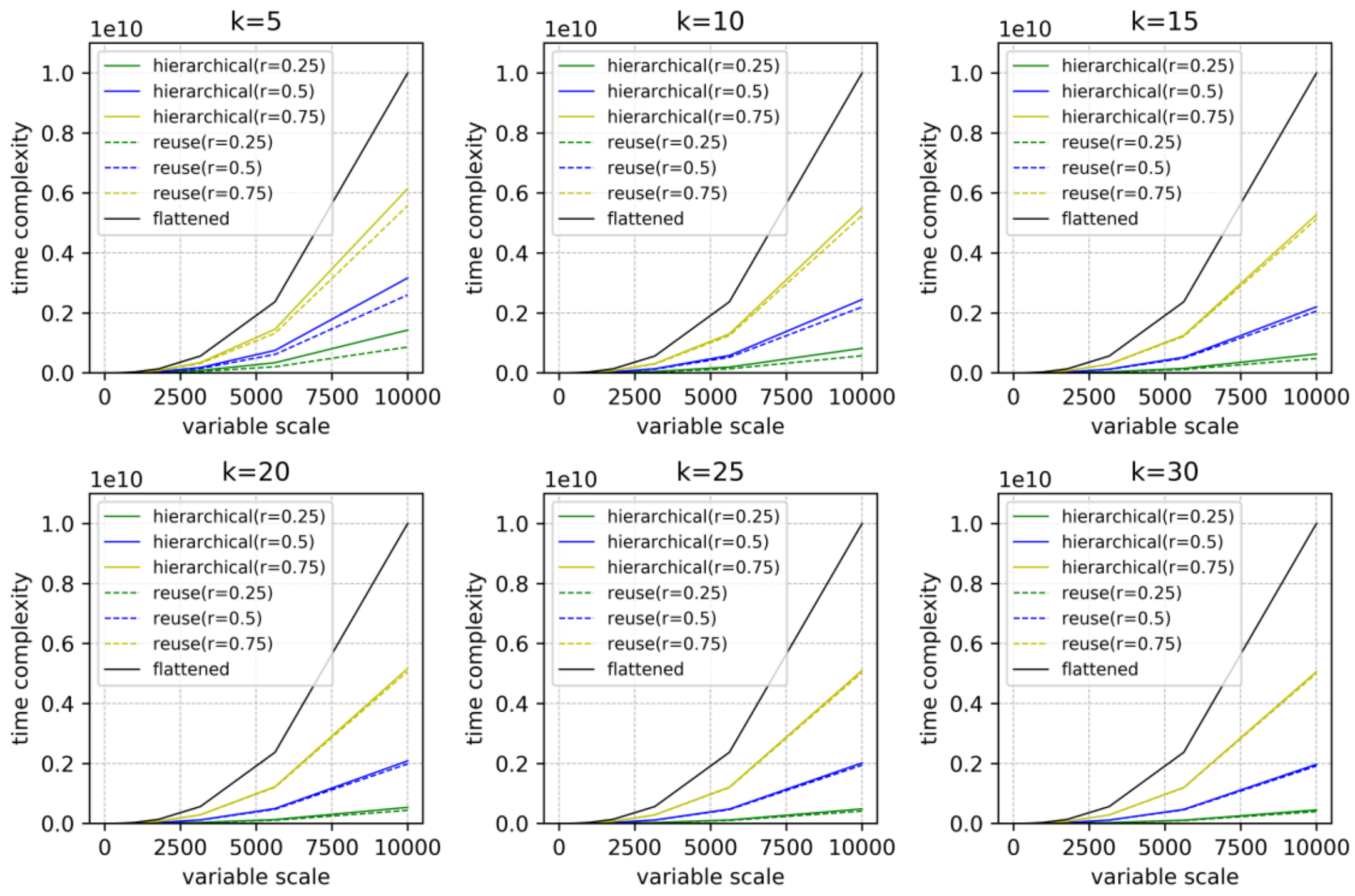

**Figure 11.** Time complexity comparison of the flattened method and the hierarchical method under different conditions.

According to the comparison in Figure 11, the following conclusions can be drawn: (1) the hierarchical structural analysis method is more efficient than the existing structural analysis method based on flattened models; (2) reusing pre-computed decomposing results of the components has the least time cost; (3) in the time complexity comparison at different values of $r$, the hierarchical structural analysis becomes more efficient as $r$ decreases; (4) at a specified variable scale $n$ and a specified under-constrained ratio $r$, the proposed method becomes more efficient as the component number $k$increases, but the increment slows down when $k$ reaches a certain degree; and (5) furthermore, the reuse of pre-computed decomposing results raises the efficiency more when $k$ becomes smaller.

For DAE models, the structural analysis will augment the equation system when searching for a maximum matching. The time complexity of this step is $O\big((2 * (n+m) + c_{\text{diff}}) * n/2\big) = O(n^3)$, where $c_{\text{diff}}$ is the times the equation is differentiated and $n$ is the number of nodes in the augmented equation system. The time complexity of decomposing a component of a DAE model is

$$C(n) \leq n^3 + (2 * c + 3) * n \tag{12}$$

The subsequent steps are based on the augmented equation system. The time complexities of constructing the dummy model and the structural analysis of the dummy model for DAE and NLAE models are similar. The difference in decomposing cost does not change our conclusion on the efficiency and the influencing factors of the efficiency.

In practice, the under-constrained ratio of the components becomes smaller as the model becomes complex. The reason for this is that the number of variables increases more quickly than the variables related to other parts. Moreover, when more reuse occurs in the modeling, the structural analysis benefits more from the hierarchical method. However, for EoMs without hierarchical structure, this method has no remarkable advantages compared with the existing methods, because the decomposition of a flat equation system has high time complexity (i.e., BLT decomposition and DM decomposition need $O(n^3)$ operations). Random decomposition of the bipartite graph forms components with a large under-constrained ratio. The efficiency of the hierarchical analysis method is the usage of sparsity from its hierarchical structure. It is hard to find an algorithm to efficiently decompose an equation system into different parts with a low under-constrained ratio.

## 6. Conclusions

This study proposed a hierarchical structural analysis method to reveal the singularity of complex EoMs. The proposed method utilizes the relationship between the singularities of a hierarchical EoM and its components. It analyzes the structural singularity of an EoM by decomposing its components, constructing a dummy model and analyzing the dummy model. The structural singularity of complex models can be analyzed by repeating this process layer by layer in their natural structure. Mathematical proofs show that the result obtained by the proposed method is equivalent to the structural analysis result based on the flattened model. Moreover, the hierarchical structural analysis method can be adaptively applied to NLAE and DAE models to realize their structural analysis. The main algorithms and application comparisons were presented for NLAE and DAE models to verify the efficiency of the proposed method. At last, the time complexity of the hierarchical structural analysis method is analyzed, and the result is compared with the time complexity of the existing structural analysis methods based on the flattened models. The comparison shows that the proposed method has better performance than the existing methods [12,31] for structural analysis of complex hierarchical EoMs, especially in the collaborative modeling environment where the components are reused, such as the crowdsourcing design platform in our previous work [41].

Compared with the existing methods, the hierarchical structural analysis method does not flatten the hierarchical EoMs. It performs the analysis layer by layer and reduces the equation scale in each analysis, thereby enabling the efficient structural analysis of complex EoMs. Based on the sparse structure of hierarchical EoMs, the proposed method avoids the defects and extra computation cost from decomposing the flattened equation system. As the comparison of the time complexity shows, the hierarchical structural analysis method has obvious advantages over the existing structural analysis methods, although the efficiency gap is affected by multiple factors. However, the application scope of the proposed method is limited in the EoMs modularly modeled in a hierarchical structure. The singularity equivalence of the dummy model and the flattened model depends on the components having no over-constrained nodes.

All algorithms in this paper have been implemented in Python to verify the efficiency of the hierarchical structural analysis method. The scripts and test cases are available on GitHub (https://github.com/wangchustcad/hierarchicalStructuralAnalysis, accessed on 09 10 2021). In the future, the algorithms should be implemented by more efficient languages, such as C++, for practice application. Moreover, the study may be extended to the hierarchical structural analysis of general EoMs in the future by finding an efficient decomposition algorithm.